\documentclass[
reprint,
nofootinbib,
superscriptaddress,
amsmath,
amssymb,
a4paper,
aps,
prd,
floatfix,
]{revtex4-2}

\usepackage{bm}
\usepackage{amsmath,amsthm, amssymb}
\usepackage{slashed}
\usepackage{dsfont}

\newcommand{\I}{\mathrm{i}}

\newcommand{\TO}{\rightarrow}

\newcommand{\diag}{\mathrm{diag}}
\newcommand{\INT}{\int d^4 x\;}
\newcommand{\g}[1]{\gamma^{#1}}
\newcommand{\PP}{\mathrm{P}}

\newcommand{\PA}{\mathrm{PA}}

\newcommand{\Tr}{\mathrm{Tr}}
\newcommand{\eq}{Eq. }
\newcommand{\su}{SU(2)_{CS}}
\newcommand{\sun}{SU(2)_{CS}^{\mathcal{P}}}
\newcommand{\sutr}{U_{CS^{\mathcal{P}}}^{\bm{\alpha}}}
\newcommand{\suno}{U_{CS}^{\bm{\alpha}}}
\newcommand{\spiu}{[(\mathds{1}+\sigma^1)\otimes\mathds{1}]}
\newcommand{\ggu}{\Gamma\,}

\begin{document}
	
	\title{A chiralspin symmetry in QCD in Minkowski space-time} 
	
	\author{Marco Catillo}

	\email{mcatillo@phys.ethz.ch}

	\affiliation{Institute for Theoretical Physics, ETH Z\"urich, 8093 Z\"urich, Switzerland}
	
	\date{\today}	
	
	\begin{abstract}
		In this paper, we look how to construct in Minkowski space-time a new type of \textit{chiralspin} group transformation of the spinor fields, similar to the one discovered by recent works of \textit{Glozman et al.} in the context of high temperature QCD and truncated studies in lattice calculations. Afterwards, we prove the invariance of free massless fermionic action under such group transformations, as well the invariance of the Hamiltonian of free massless fermions.
		At the end, the possible presence of a symmetry driven by such new \textit{chiralspin} group at high temperature QCD, also at non zero chemical potential, is discussed. 
	\end{abstract}

\maketitle

\section{Introduction}\label{sec:introduction}
	
	In recent works, the emergence of an unexpected symmetry in lattice QCD simulations has been observed. In particular at high temperature QCD \cite{Rohrhofer:2017grg,Rohrhofer:2019qwq,Rohrhofer:2019qal,Glozman:2021jlk}, right above the chiral phase transition $T>T_c$, but for $T \lesssim 3 T_c $, and in truncated studies (see Refs.  \cite{Denissenya:2014poa,Denissenya:2014ywa,Denissenya:2015mqa} for information on these peculiar works). 
	More specifically, in truncated studies a large degeneration of hadron masses has been discovered.
	The symmetry which corresponds to such degeneracy has been explained by the group transformation $\su$ (or in words \textit{chiralspin} group) of the quark fields, first introduced in \cite{Denissenya:2014poa,Denissenya:2014ywa,Denissenya:2015mqa}, 
	and that contains the axial group $U(1)_A$ as subgroup. 
	However, as we have studied in \cite{Catillo:2021rrq}, the mass degeneracy could also be explained, in the Euclidean space-time, by the group transformation which we have denoted with $\sun$, that is defined in a slight different manner from $\su$, but still has $U(1)_A$ as subgroup. 
	In fact, the two group transformations induce the same transformation in hadron correlators $\langle \mathcal{O}(y)\bar{\mathcal{O}}(x)\rangle$ calculated at fixed reference frame with $x = (\bm{0},x_4)$ and $y=(\bm{0},y_4)$, from which we can still extract the hadron masses, since at large $\mathcal{T} = y_4 - x_4$, we have $\langle \mathcal{O}(y)\bar{\mathcal{O}}(x)\rangle \sim \exp(-m\,\mathcal{T})$, 
	with $m$ the hadron mass associated to such correlator. 
	Moreover, we have seen that, while $\su$ is not a symmetry of the free fermionic action, which makes it not compatible with the possibility of deconfinement at extremely high T, $\sun$ is instead a symmetry of the free fermionic action,  which makes it more suitable to check its presence at $T\gg T_c$, where QCD is supposed to approach at an almost free theory. 
	
	The work done in Ref. \cite{Catillo:2021rrq} has been considered in Euclidean space-time. 
	Here, we see that we can define the $\sun$ either in Minkowskian and also prove that it leaves the fermionic action invariant, repeating the same argumentation of \cite{Catillo:2021rrq} (see in section \ref{sec:chiralspin2} of this paper). 
	For doing so, we need to define a $U(1)$ group starting simply from the parity operator (look section \ref{sec:chiralspin}). 
	Beside this, we also prove the invariance of the Hamiltonian of free massless fermions under $\sun$, giving how the operators of creation and annihilation of quarks and antiquarks (but in general fermions and anti-fermions) transform under $\sun$ (in section \ref{sec:ham}). 
	We also briefly discuss what happens when a gauge interaction term is added in the theory. 
	Finally, we repeat in Minkowskian the argument made in Ref. \cite{Catillo:2021rrq}, regarding the presence of $\sun$ at high T. 
	Moreover we will see that a possible chemical potential term in the action is $\sun$-invariant. 
	Therefore we expect that if $\sun$ would be present at high T and zero chemical potential, i.e. $\mu = 0$, it will be present also at $\mu\neq 0 $ (section \ref{sec:correlators}). 
	Finally we summarize the main points of this paper in section \ref{sec:summary}. 
	
	We remark that from section \ref{sec:chiralspin} to \ref{sec:ham} (and including all appendices), everything is kept general, and the reader can assume that we are considering a theory with whatever gauge group $\mathcal{G}$. 
	The section \ref{sec:correlators} is instead specific for QCD, where $\mathcal{G} = SU(3)$, because it is in connection with the lattice results of Refs. \cite{Denissenya:2014poa,Denissenya:2014ywa,Denissenya:2015mqa,Rohrhofer:2017grg,Rohrhofer:2019qwq,Rohrhofer:2019qal,Glozman:2021jlk}.
	
	\section{From parity to a $U(1)$ group}\label{sec:chiralspin}
	
	For a spinor field, the parity transformation is defined as $\psi(x)\to P\psi(x) P^{\dagger} = \g{0} \psi(\mathcal{P}x)$, with $\mathcal{P}=\diag(1,-1,-1,-1)$ and $P$ the parity operator with properties $P=P^{\dagger}$ and $P^2 = \mathds{1}$. 
	The application of two times this transformation gives back again the same spinor field, because $P^2\psi(x) P^{\dagger\,2} = \g{0}P\psi(\mathcal{P}x)P^{\dagger} = \g{0}\g{0}\psi(\mathcal{P}^2x) = \psi(x)$, since $\mathcal{P}^2 = \mathcal{I}$ and $(\g{0})^2 = \mathds{1}$ 
	(see Eq. (\ref{eq:notgamma}) for the representation used for the gamma matrices in this paper). 
	Therefore, for $n$ applications of parity, we have $P^n\psi(x) P^{\dagger\,n} = (\g{0})^n\psi(\mathcal{P}^n x)$, which is $\psi(x)$ for $n$ even and $\g{0}\psi(\mathcal{P}x)$ for $n$ odd. 
	Exploiting this fact we can define the following spinor transformation: 
	
	\begin{equation}
		\begin{split}
		U(1)_{\PP}:\;&\psi(x)\to\psi(x)^{U_{\PP}^{\alpha}} 
		\equiv\sum_{n=0}^{\infty}\frac{(\I\alpha)^n}{n!}P^n\psi(x) P^{\dagger\,n}\\
		&= \cos(\alpha)\,\psi(x) +\I\sin(\alpha)\, \g{0}\psi(\mathcal{P}x),
		\end{split}
		\label{eq:uid}
	\end{equation}
	
	\noindent
	where $\alpha$ is some global parameter. Eq. (\ref{eq:uid}) is basically an exponentiation of the parity transformation.
	Such transformation is of course linear, 
	in the sense that  taking two generic fields $\psi_1(x)$ and $\psi_2(x)$, then $(\psi_1(x) + \psi_2 (x))^{U_{\PP}^{\alpha}}=\psi_1(x)^{U_{\PP}^{\alpha}} + \psi_2(x)^{U_{\PP}^{\alpha}}$, which simply comes from the definition (\ref{eq:uid}) and observing that $P^n (\psi_1(x) + \psi_2 (x))P^{\dagger\,n} = P^n \psi_1(x)P^{\dagger\,n} + P^n\psi_2 (x)P^{\dagger\,n}$. 
	
	The name that we have chosen for such transformations in (\ref{eq:uid}), comes from the fact that they form a $U(1)$ group. 
	In order to show this point directly, we introduce a bit of notation. 
		We construct two fields $\psi_{\pm}(x) = \frac{1}{2}(\psi(x) \pm\psi(\mathcal{P}x))$, that we call ``parity partners'' and satisfying the properties: $\g{0}\psi_{\pm}(x) = \pm P\psi_{\pm}(x)P^{\dagger}$ and $\psi_{\pm}(\mathcal{P}x) = \pm\psi_{\pm}(x)$. Afterwards we define the 2-component field 
	
	\begin{equation}
	\Psi(x) =\left(
	\begin{matrix}
	\psi_{+}(x)\\
	\psi_{-}(x)
	\end{matrix}
	\right).
	\label{eq:psi}
	\end{equation}
	
	\noindent
	Now, we can transform $\psi_{+}(x)$ and $\psi_{-}(x)$ via $U(1)_{\PP}$ separately, 
	and use the linear property previously mentioned, 
	obtaining that 
	$\psi_{\pm}(x)^{U_{\PP}^{\alpha}} = \frac{1}{2}(\psi(x)^{U_{\PP}^{\alpha}}\pm\psi(\mathcal{P}x)^{U_{\PP}^{\alpha}})$. 
	
	Therefore $\Psi(x)$ transforms as 
	
	\begin{equation}
	\begin{split}
	U(1)_{\PP}:\,&\Psi(x)\to\Psi(x)^{U_{\PP}^{\alpha}} = \left(
	\begin{matrix}
	\psi_{+}(x)^{U_{\PP}^{\alpha}}\\
	\psi_{-}(x)^{U_{\PP}^{\alpha}}\\
	\end{matrix}
	\right)\\
	&=
	\left(
	\begin{matrix}
	e^{\I\alpha \g{0}} \psi_{+}(x)\\
	e^{-\I\alpha \g{0}} \psi_{-}(x)\\
	\end{matrix}
	\right)
	=e^{\I\alpha(\sigma^3\otimes\g{0})}\Psi(x),
	\end{split}
	\label{eq:psigen}
	\end{equation} 
	
	\noindent
	where $\sigma^3$ is the 3rd Pauli matrix.  
	From (\ref{eq:psigen}) is evident that $U(1)_{\PP}$ transformations, acting on $\Psi(x)$, form a $U(1)$ group with generator $\sigma^3\otimes \g{0}$, 
	which is hermitian and traceless, because $(\sigma^3\otimes\g{0})^{\dagger} = (\sigma^3)^{\dagger}\otimes(\g{0})^{\dagger} = \sigma^3\otimes\g{0}$ 
	and $\Tr(\sigma^3\otimes\g{0}) = \Tr(\sigma^3)\Tr(\g{0}) = 0$. 
	
	We can prove now that the $U(1)_{\PP}$ group leaves invariant the action of free massive fermions.
	For checking this, we can see the transformation of the conjugate field $\bar{\psi}(x)$, which from (\ref{eq:uid}) is just $\bar{\psi}(x)^{U_{\PP}^{\alpha}} = (\psi(x)^{U_{\PP}^{\alpha}})^{\dagger}\g{0}$. 
	Therefore the free fermion action 
	
	\begin{equation}
		S_{F}(\psi,\bar{\psi}) = \int d^4 x\, \bar{\psi}(x)\, (\I\g{\mu}\partial^x_{\mu}-m)\,\psi(x).
		\label{eq:sf}
	\end{equation}
	
	\noindent
	with $\partial^x_{\mu} = \frac{\partial}{\partial x^{\mu}}$ and where the integration has to be intended over the whole space-time, transforms as
	
	\begin{equation}
	\begin{split}
	&S_F(\psi^{U_{\PP}^{\alpha}},\bar{\psi}^{U_{\PP}^{\alpha}})=
	\INT \bar{\psi}(x)^{U_{\PP}^{\alpha}}(\I\g{\mu}\partial_{\mu}^x-m)\psi(x)^{U_{\PP}^{\alpha}}\\
	&=\cos(\alpha)^2 \INT \bar{\psi}(x)(\I\g{\mu}\partial_{\mu}^x-m)\psi(x)\\
	&+\I\sin(\alpha)\cos(\alpha)\INT\bar{\psi}(x)(\I\g{\mu}\partial_{\mu}^x-m)\g{0}\psi(\mathcal{P}x)\\
	&-\I\sin(\alpha)\cos(\alpha)\INT\bar{\psi}(\mathcal{P}x)\g{0}(\I\g{\mu}\partial_{\mu}^x-m)\psi(x)\\
	&+\sin(\alpha)^2\INT\bar{\psi}(\mathcal{P}x)\g{0}(\I\g{\mu}\partial_{\mu}^x-m)\g{0}\psi(\mathcal{P}x)\\
	&=\cos(\alpha)^2\INT\bar{\psi}(x)(\I\g{\mu}\partial_{\mu}^x-m)\psi(x)\\
	&+\I\sin(\alpha)\cos(\alpha)\INT \bar{\psi}(x)(\I\g{\mu}\partial_{\mu}^x-m)\g{0} \psi(\mathcal{P}x)\\
	&-\I\sin(\alpha)\cos(\alpha)\INT \bar{\psi}(\mathcal{P}x)(\I\g{\mu}\partial_{\mu}^{\mathcal{P}x}-m)\g{0} \psi(x)\\
	&+\sin(\alpha)^2\INT\bar{\psi}(\mathcal{P}x)(\I\g{\mu}\partial_{\mu}^{\mathcal{P}x}-m)\psi(\mathcal{P}x)\\
	&=\INT \bar{\psi}(x)(\I\g{\mu}\partial_{\mu}^{x}-m)\psi(x)= S_{F}(\psi,\bar{\psi}),
	\end{split}
	\label{eq:action}
	\end{equation}
	
	\noindent
	where in the 2nd equality we just expanded using (\ref{eq:uid}),
	in the 3rd equality for the 2nd and 3rd term, 
	we used the property $\g{0}\g{\mu}\g{0} = \g{\nu}\;\mathcal{P}_{\nu}^{\;\mu}$ and denoted $\mathcal{P}_{\mu}^{\;\nu}\,\partial_{\nu}^{\mathcal{P}x} = \partial_{\mu}^x$.
	Moreover we have changed the variable $x\to\mathcal{P}x$,
	and used that the Jacobian $\vert\det(\mathcal{P})\vert=1$. After this, we see that the terms proportional to $\cos(\alpha)^2$ and $\sin(\alpha)^2$ sum up to $1$, while the ones proportional to $\I\sin(\alpha)\cos(\alpha)$ sum up to $0$.
	As it is clear from the previous equation, $U(1)_{\PP}$ transformations leave invariant $S_F$. 
	
	However, the introduction of a gauge field interaction
	like
	\begin{equation}
		S_I (\psi,\bar{\psi},A) = g\INT\bar{\psi}(x)\g{\mu}A_{\mu}(x)\psi(x),
	\end{equation}
	 \noindent
	 breaks explicitly $U(1)_{\PP}$.
	This was already described in \cite{Catillo:2021rrq} in Euclidean space-time,  but the same argument can be trivially translated here in Minkowskian using the transformations (\ref{eq:uid}), and it comes by the following lines,

	\begin{equation}
	\begin{split}
	&S_I (\psi^{U_{\PP}^{\alpha}},\bar{\psi}^{U_{\PP}^{\alpha}},A)/g= \INT\, \bar{\psi}(x)^{U_{\PP}^{\alpha}}\g{\mu}A_{\mu}(x)\psi(x)^{U_{\PP}^{\alpha}}\\
	&= \cos(\alpha)^2 \INT\, \bar{\psi}(x)\,\g{\mu}A_{\mu}(x)\,\psi(x)\\
	&+\I\sin(\alpha)\cos(\alpha)\INT\,\bar{\psi}(x)\,\g{\mu}\g{0}A_{\mu}(x)\,\psi(\mathcal{P}x)\\
	&-\I\sin(\alpha)\cos(\alpha)\INT\,\bar{\psi}(\mathcal{P}x)\,\g{0}\g{\mu}A_{\mu}(x)\,\psi(x)\\
	&+\sin(\alpha)^2 \INT\,\bar{\psi}(\mathcal{P}x)\,\g{0}\g{\mu}\g{0}A_{\mu}(x)\,\psi(\mathcal{P}x)\\
	&= \cos(\alpha)^2 \INT\, \bar{\psi}(x)\,\g{\mu}A_{\mu}(x)\,\psi(x)\\
	&+\I\sin(\alpha)\cos(\alpha)\INT\,\bar{\psi}(x)\,\g{\mu}\g{0}A_{\mu}(x)\,\psi(\mathcal{P}x)\\
	&-\I\sin(\alpha)\cos(\alpha)\INT\,\bar{\psi}(x)\,\mathcal{P}_{\nu}^{\;\mu}\g{\nu}\g{0}A_{\mu}(\mathcal{P}x)\,\psi(\mathcal{P}x)\\
	&+\sin(\alpha)^2 \INT\,\bar{\psi}(x)\,\g{\nu}\mathcal{P}_{\nu}^{\;\mu}A_{\mu}(\mathcal{P}x)\,\psi(x)\\
	&=\INT \left[\bar{\psi}(x)\, \g{\mu}(\cos(\alpha)^2 A_{\mu}(x) + \sin(\alpha)^2 A_{\mu}^P(x))\,\psi(x) \right.\\
	&+\left.\I\sin(\alpha)\cos(\alpha)\,\bar{\psi}(x)\, \g{\mu}(A_{\mu}(x) - A_{\mu}^P(x))\,\g{0}\,\psi(\mathcal{P}x)\right], 
	\end{split}
	\label{eq:actioninte}
	\end{equation}
	
	\noindent
	where we used the same procedure as in Eq. (\ref{eq:action}) and we defined  $A_{\nu}^P(x)\equiv\mathcal{P}_{\nu}^{\;\mu}A_{\mu}(\mathcal{P}x) = P A_{\nu}(x) P^{\dagger}$.
	As it is clear from (\ref{eq:actioninte}), for generic values of $\alpha$, $S_I (\psi^{U_{\PP}^{\alpha}},\bar{\psi}^{U_{\PP}^{\alpha}},A) \neq S_I (\psi,\bar{\psi},A)$, because in general $A_{\mu}^P(x)\neq A_{\mu}(x)$, 
	and a $U(1)_{\PP}$ transformation mixes both these fields. 
	Therefore the interaction term breaks $U(1)_{\PP}$. 
	However if we restrict to particular values of $\alpha$, we can obtain the invariance of $S_I$. In particular we recognize two cases:
	
	\begin{itemize}
		\item $\alpha = \pi k$ with $k=0,1,2,...$ $\Rightarrow U(1)_{\PP}$ reduces to the group $Z_2 \subset U(1)_{\PP}$ and of course $S_F + S_I$ is $Z_2$-invariant.
		\item $\alpha = \pi k+ (\pi/2)$ with $k=0,1,2,...$ $\Rightarrow$ In this case if we perform also a parity transformation of the gauge field $A_{\mu}(x)\to A_{\mu}^{P}(x)$, we can obtain the invariance of the interaction term, namely $S_I (\psi^{U_{\PP}^{\alpha}},\bar{\psi}^{U_{\PP}^{\alpha}},A^P) = S_I (\psi,\bar{\psi},A)$. 
		In fact, as it is clear from Eq. (\ref{eq:uid}), $U(1)_{\PP}$  transformations reduce to Parity $\times\,\I  Z_2 \in U(1)_{\PP}$ transformations, 
		where for Parity $\times\, \I Z_2 $ we mean for example the transformation: $\psi(x) \to z\,\I (P\psi(x)P^{\dagger})$, with $z\in Z_2$.
	\end{itemize}
	
	Otherwise a sufficient condition for the $U(1)_{\PP}$ invariance of $S_I (\psi,\bar{\psi},A)$, can be obtained restricting ourself to gauge configurations such that 
	
	\begin{equation}
		A_{\mu}^{P}(x)=A_{\mu}(x),
		\label{eq:gauge}
	\end{equation} 
	
	\noindent
	which means $A_{0}(\mathcal{P}x)=A_{0}(x)$ and $A_{i}(\mathcal{P}x)=-A_{i}(\mathcal{P}x)$. 

	\section{Towards a new chiralspin group}\label{sec:chiralspin2}
	
	Beside $U(1)_{\PP}$, the other ingredient that we need for constructing our new \textit{chiralspin} group is to derive the $U(1)_A$ transformations for the field given in (\ref{eq:psi})  
	$U(1)_A$ transformations are defined on $\psi$ as 
	$\psi(x)\to \psi(x)^{U_{A}^{\alpha}} = \exp(-\I\alpha\g{5})\psi(x)$ and changing the variable $x\to\mathcal{P}x$, we have the same, i.e.  $\psi(\mathcal{P}x)\to \psi(\mathcal{P}x)^{U_{A}^{\alpha}} = \exp(-\I\alpha\g{5})\psi(\mathcal{P}x)$. 
	Therefore this translates to $\psi_{+}$ and $\psi_{-}$ constructing 
	$\psi_{\pm}^{U_A^{\alpha}}(x) =\frac{1}{2}(\psi(x)^{U_{A}^{\alpha}}\pm   \psi(\mathcal{P}x)^{U_{A}^{\alpha}})$, so that $\psi_{\pm}^{U_A^{\alpha}}(x) = \exp(-\I\alpha\g{5})\psi_{\pm}(x)$. 
	Hence from the definition in (\ref{eq:psi}), we have 
	
	\begin{equation}
	\begin{split}
	U(1)_{A}:\,&\Psi(x)\to\Psi(x)^{U_{A}^{\alpha}} = \left(
	\begin{matrix}
	\psi_{+}(x)^{U_{A}^{\alpha}}\\
	\psi_{-}(x)^{U_{A}^{\alpha}}\\
	\end{matrix}
	\right)\\
	&=
	\left(
	\begin{matrix}
	e^{-\I\alpha \g{5}} \psi_{+}(x)\\
	e^{-\I\alpha \g{5}} \psi_{-}(x)\\
	\end{matrix}
	\right)
	=e^{\I\alpha(-\mathds{1}\otimes\g{5})}\Psi(x).
	\end{split}
	\label{eq:psigen2}
	\end{equation}
	
	\noindent
	The generator of $U(1)_A$ for the field $\Psi$ is therefore $-\mathds{1}\otimes\g{5}$, which is traceless and hermitian.

	\subsection{New chiralspin group definition}
	
	Taking now the generators of the groups $U(1)_A$ and $U(1)_{\PP}$, we rename them as  $\Sigma_1^{\mathcal{P}} = \sigma^3\otimes\g{0}$, 
	and $\Sigma_3^{\mathcal{P}} = -\mathds{1}\otimes\g{5}$ and we define the third matrix $\Sigma_2^{\mathcal{P}}  = \I\Sigma_1^{\mathcal{P}} \Sigma_3^{\mathcal{P}}= \sigma^3\otimes\I\g{5}\g{0}$, which is still traceless and hermitian, since $\Tr(\sigma^3\otimes\I\g{5}\g{0})= \Tr(\sigma^3)\Tr(\I\g{5}\g{0})=0$ and $(\sigma^3\otimes\I\g{5}\g{0})^{\dagger} = (\sigma^{3})^{\dagger}\otimes(\I\g{5}\g{0})^{\dagger} = \sigma^{3}\otimes\I\g{5}\g{0}$ (see Eq. (\ref{eq:notgamma})). 
	Now the set of $\Sigma_n^{\mathcal{P}}$s, which are all traceless and hermitian, 
	verify the property: $[\Sigma_i^{\mathcal{P}},\Sigma_j^{\mathcal{P}}]=2\I\epsilon_{ijk}\Sigma_k^{\mathcal{P}}$. 
	Hence they are generators of an $su(2)$ algebra. 
	We call the Lie group generated by the $\Sigma_n^{\mathcal{P}}$s as $\sun$.  
	The $\sun$ group transformations on the field $\Psi$ in (\ref{eq:uid}) are given by
	
	\begin{equation}
	\begin{split}
	\sun\,:
	&\Psi(x)\to\Psi(x)^{\sutr} =\sutr \Psi(x),\\ 
	&\sutr = e^{\I\alpha_n\Sigma_n^{\mathcal{P}}}\in\sun,
	\end{split}
	\label{eq:su2csnew}
	\end{equation}
	
	\noindent
	from which for the proper choice of the global vector $\bm{\alpha} =(\alpha_1,\alpha_2,\alpha_3)$, we can get the group transformations of $U(1)_{\PP}$ and $U(1)_A$ in (\ref{eq:psigen}) and (\ref{eq:psigen2}) respectively. 
	This means that $U(1)_{\PP},U(1)_A \subset \sun$. 
	From the transformations (\ref{eq:su2csnew}) we can get how $\psi$ (and consequently $\bar{\psi} = \psi^{\dagger}\g{0}$) transforms, just inverting the definition of $\psi_{\pm}$ in terms of $\psi$. 
	
	Let us see now how to do it. 
	First of all we recall an important feature which will be also useful  later on. 
	As it is well-known every element of a $SU(2)$ group can be written as product of three $U(1)$ matrices, which are subgroups of $SU(2)$.
	More precisely, if $\sutr \in\sun$, it can be always written as $\sutr = U_{\PP}^{\beta_1}U_A^{\beta_2}U_{\PP}^{\beta_3}$, 
	where the $\beta_i$s are the three Euler angles, while $U_{\PP}^{\beta_{1,3}} = \exp(\I\beta_{1,3}(\sigma^3\otimes\g{0}))\in U(1)_{\PP}$ and $U_A^{\beta_2} = \exp(\I\beta_2(-\mathds{1}\otimes\g{5}))\in U(1)_A$. 
	Therefore $\Psi(x)^{\sutr} =\sutr \Psi(x) = U_{\PP}^{\beta_1}U_A^{\beta_2}U_{\PP}^{\beta_3}\Psi(x) \Rightarrow 
	\Psi(x)^{\sutr} = ((\Psi(x)^{U_{\PP}^{\beta_3}})^{U_{A}^{\beta_2}})^{U_{\PP}^{\beta_1}}$. 
	This means that the two subgroups $U(1)_{\PP}$ and $U(1)_A$ can give whatever matrix of $\sun$.
	Now from $\Psi(x)^{U_{\PP}^{\beta_3}}$ we get the transformation of $\psi(x)$, just using Eq. (\ref{eq:psigen}), namely $\Psi(x)^{U_{\PP}^{\beta_3}}\Rightarrow\psi(x)^{U_{\PP}^{\beta_3}} = \psi_{+}(x)^{U_{\PP}^{\beta_3}}+ \psi_{-}(x)^{U_{\PP}^{\beta_3}}$. 
	Moreover, calling $\psi'(x) = \psi(x)^{U_{\PP}^{\beta_3}}$, 
	from Eq. (\ref{eq:psigen2}), we get $(\Psi(x)^{U_{\PP}^{\beta_3}})^{U_{A}^{\beta_2}}\Rightarrow \psi'(x) ^{U_{A}^{\beta_2}} = 
	\psi'_{+}(x)^{U_{A}^{\beta_2}} + \psi'_{-}(x)^{U_{A}^{\beta_2}}$. 
	Finally, naming $\psi''(x) = \psi(x)^{U_{A}^{\beta_2}}$, again from  (\ref{eq:psigen}), we have
	$((\Psi(x)^{U_{\PP}^{\beta_3}})^{U_{A}^{\beta_2}})^{U_{\PP}^{\beta_1}}\Rightarrow\psi''(x)^{U_{\PP}^{\beta_1}} =\psi''_{+}(x)^{U_{\PP}^{\beta_1}} + \psi''_{-}(x)^{U_{\PP}^{\beta_1}} $. 
	Therefore the $\sun$ transformation of $\psi$ is given by $\psi(x)^{\sutr} \equiv \psi''(x)^{U_{\PP}^{\beta_1}}$, hence 
	
	\begin{equation}
		\psi(x)^{\sutr} = ((\psi(x)^{U_{\PP}^{\beta_3}})^{U_{A}^{\beta_2}})^{U_{\PP}^{\beta_1}},
		\label{eq:eul}
	\end{equation}
	
	\noindent
	and its full expression using (\ref{eq:uid}) is
	
	\begin{equation}
	\begin{split}
				\psi(x)^{\sutr}  &= 
		\cos(\beta_1)\left[\exp(-\I\beta_2\g{5})(\cos(\beta_3)\psi(x)\right.\\
		&\left.+\I\sin(\beta_3)\g{0}\psi(\mathcal{P}x))\right]\\
		&+\I\sin(\beta_1)\,\g{0}\left[\exp(-\I\beta_2\g{5})(\cos(\beta_3)\psi(\mathcal{P}x)\right.\\
		&\left. + \I\sin(\beta_3)\g{0}\psi(x))\right],
	\end{split}
	\label{eq:psisu2cs}
	\end{equation}
	
	\noindent
	which is what we wanted to get. 
	
	A particular case is when in (\ref{eq:su2csnew}), we set  $(\alpha_1,\alpha_2,\alpha_3) = (0,\alpha,0)\equiv\bar{\bm{\alpha}}$. 
	In this situation, we obtain another $U(1)$ subgroup of $\sun$, which we call $U(1)_{\PA}$ and its generator is therefore $\Sigma_2^{\mathcal{P}} = \sigma^3\otimes\I\g{5}\g{0}$. 
	Now whatever element $U_{\PA}^{\alpha}\equiv \exp(\I\alpha\Sigma_2^{\mathcal{P}})\in U(1)_{\PA}$ can be written always using the Euler angles as $U_{\PA}^{\alpha} = U_{\PP}^{\pi/4} U_{A}^{\alpha}U_{\PP}^{-\pi/4}$, that it is easy to verify from (\ref{eq:su2csnew}) and (\ref{eq:psisu2cs}). 
	This means that in the Euler decomposition we have $(\beta_1,\beta_2,\beta_3) = (\pi/4,\alpha,-\pi/4)$. Substituting such values in Eq. (\ref{eq:psisu2cs}) and defining 
	$\psi(x)^{U_{\PA}^{\alpha}}\equiv \psi(x)^{U_{CS^{\mathcal{P}}}^{\bm{\bar{\alpha}}}} $, we 
	obtain 
	
	\begin{equation}
	\begin{split}
		U(1)_{\PA}:\,&\psi(x)\to\psi(x)^{U_{\PA}^{\alpha}}\\
		& = 
	\cos(\alpha)\,\psi(x) + \I\sin(\alpha) (\I\g{5}\g{0})\,\psi(\mathcal{P}x),
		\end{split}
	\label{eq:u1pa}
	\end{equation}
	
	\noindent
	which is the $U(1)_{\PA}$ group transformation of $\psi$. 
	As we can see it is similar to the $U(1)_{\PA}$ transformations defined in Ref. \cite{Catillo:2021rrq} for the Euclidean case.
	
	We conclude saying that the group $\sun$ as defined by Eq. (\ref{eq:su2csnew}) differently from $\su$ in Ref. \cite{Denissenya:2015mqa}, looks as a rotation in the space of the ``parity partners'' $\psi_{+}(x)$ and $\psi_{-}(x)$ (which is similar, but not the same, of what we did in Ref. \cite{Catillo:2018cyv} for baryon parity doublets).

	\subsection{Consequences on the fermionic action}
	
	From how $\sun$ is defined in Eq. (\ref{eq:su2csnew}) we can obtain some consequences on the invariance of the fermionic action, in particular
	
	\begin{enumerate}
		\item $S_F (\psi,\bar{\psi})$ at $m=0$ is $SU(2)_{CS}^{\mathcal{P}}$-invariant, 
		\item the mass term of $S_F (\psi,\bar{\psi})$ breaks explicitly $U(1)_{\PA}$ and moreover a gauge interaction in the action is not $U(1)_{\PA}$-invariant,
		\item a gauge interaction breaks $\sun$. However if we restrict to gauge fields satisfying the relation given in (\ref{eq:gauge}), then $S_I (\psi,\bar{\psi},A)$ is $\sun$-invariant.
	\end{enumerate}

	For proving the first statement, we use the decomposition of $\psi(x)^{\sutr}$ in Eq. (\ref{eq:eul})
	 and the $U(1)_{\PP}$-invariance of $S_F (\psi,\bar{\psi})$ in Eq. (\ref{eq:action}). 
	Hence we get $S_F (\psi^{\sutr},\bar{\psi}^{\sutr}) = S_F ((\psi^{U_{\PP}^{\beta_3}})^{U_{A}^{\beta_2}},
	(\bar{\psi}^{U_{\PP}^{\beta_3}})^{U_{A}^{\beta_2}})$. 
	 Now for $m=0$, $U(1)_A$ transformations leave invariant $S_F$, hence 
	$S_F ((\psi(x)^{U_{\PP}^{\beta_3}})^{U_A^{\beta_2}},(\bar{\psi}(x)^{U_{\PP}^{\beta_3}})^{U_A^{\beta_2}})\vert_{m=0} = S_F (\psi(x)^{U_{\PP}^{\beta_3}},\bar{\psi}(x)^{U_{\PP}^{\beta_3}})\vert_{m=0}$.
	Finally again from (\ref{eq:action}), we get $S_F (\psi(x)^{U_{\PP}^{\beta_3}},\bar{\psi}(x)^{U_{\PP}^{\beta_3}})\vert_{m=0} = S_F(\psi,\bar{\psi})\vert_{m=0}$. Therefore we obtain the $\sun$ invariance  $S_F(\psi^{\sutr},\bar{\psi}^{\sutr})\vert_{m=0}=S_F(\psi,\bar{\psi})\vert_{m=0}$. 

	The second statement is proved as follow. 
	First of all 
	$S_F (\psi^{U_{\PA}^{\alpha}},\bar{\psi}^{U_{\PA}^{\alpha}})\equiv 
	S_F (((\psi^{U_{\PP}^{-\pi/4}})^{U_{A}^{\alpha}})^{U_{\PP}^{\pi/4}},((\bar{\psi}^{U_{\PP}^{-\pi/4}})^{U_{A}^{\alpha}})^{U_{\PP}^{\pi/4}})$, 
	where we used the Euler decomposition of the previous section.
	From (\ref{eq:action}), we have $S_F (\psi^{U_{\PA}^{\alpha}},\bar{\psi}^{U_{\PA}^{\alpha}}) = S_F ((\psi^{U_{\PP}^{-\pi/4}})^{U_{A}^{\alpha}},(\bar{\psi}^{U_{\PP}^{-\pi/4}})^{U_{A}^{\alpha}})$. 
	However we already know that $U(1)_A$ is broken by the mass term. 
	Therefore $S_F ((\psi^{U_{\PP}^{-\pi/4}})^{U_{A}^{\alpha}},(\bar{\psi}^{U_{\PP}^{-\pi/4}})^{U_{A}^{\alpha}})\neq S_F (\psi^{U_{\PP}^{-\pi/4}},\bar{\psi}^{U_{\PP}^{-\pi/4}})$ 
	and we know that, from Eq. (\ref{eq:action}), $S_F (\psi^{U_{\PP}^{-\pi/4}},\bar{\psi}^{U_{\PP}^{-\pi/4}}) =S_F(\psi,\bar{\psi})$. 
	Consequently $S_F (\psi^{U_{\PA}^{\alpha}},\bar{\psi}^{U_{\PA}^{\alpha}}) \neq S_F(\psi,\bar{\psi})$ for $m\neq 0$. 
	Therefore the mass term breaks $U(1)_{\PA}$.
	The second part of the statement is similar but observing another possible Euler decomposition $U_{\PA}^{\alpha} = U_{A}^{-\pi/4}U_{\PP}^{\alpha}U_{A}^{\pi/4}$, valid for every $\alpha$. 
	Thus we have $S_I (\psi^{U_{\PA}^{\alpha}},\bar{\psi}^{U_{\PA}^{\alpha}},A)\equiv S_I (((\psi^{U_{A}^{\pi/4}})^{U_{\PP}^{\alpha}})^{U_{A}^{-\pi/4}},((\bar{\psi}^{U_{A}^{\pi/4}})^{U_{\PP}^{\alpha}})^{U_{A}^{-\pi/4}},A)$. 
	However we already know that $S_I$ is $U(1)_{A}$-invariant, 
	therefore $S_I (\psi^{U_{\PA}^{\alpha}},\bar{\psi}^{U_{\PA}^{\alpha}},A) = 
	S_I ((\psi^{U_{A}^{\pi/4}})^{U_{\PP}^{\alpha}},(\bar{\psi}^{U_{A}^{\pi/4}})^{U_{\PP}^{\alpha}},A)$ and $S_I (\psi^{U_{A}^{\pi/4}},\bar{\psi}^{U_{A}^{\pi/4}},A) = S_I (\psi,\bar{\psi},A)$. 
	Nevertheless, from (\ref{eq:actioninte}), $S_I ((\psi^{U_{A}^{\pi/4}})^{U_{\PP}^{\alpha}},(\bar{\psi}^{U_{A}^{\pi/4}})^{U_{\PP}^{\alpha}},A) \neq S_I (\psi^{U_{A}^{\pi/4}},\bar{\psi}^{U_{A}^{\pi/4}},A)$. 
	Hence $S_I (\psi^{U_{\PA}^{\alpha}},\bar{\psi}^{U_{\PA}^{\alpha}},A)\neq S_I (\psi,\bar{\psi},A)$. 
	This means that the breaking of $S_I$ under $U(1)_{\PP}$ leads to the breaking of $U(1)_{\PA}$.
	A direct proof of both parts of this second statement involving directly the $U(1)_{\PA}$ transformations (\ref{eq:u1pa}) is given in Appendix \ref{app:b}. 
	
	The last statement comes from the observation that the action $S_I$ which involves the interaction between fermions and gauge field is in general not invariant under $U(1)_{\PP}$ transformations, as it is evident from (\ref{eq:actioninte}).
	Therefore $\sun$, which has $U(1)_{\PP}$ as subgroup, is not a symmetry of $S_I (\psi,\bar{\psi},A)$ in the general case. 
	Nevertheless, if we restrict to gauge fields with the property given in Eq. (\ref{eq:gauge}), then we have $S_I (\psi^{U_{\PP}^{\beta_1}},\bar{\psi}^{U_{\PP}^{\beta_1}},A) = S_I (\psi,\bar{\psi},A)$. Therefore using the decomposition in (\ref{eq:eul}), we obtain $S_I (\psi^{\sutr},\bar{\psi}^{\sutr},A) =   S_I ((\psi^{U_{\PP}^{\beta_3}})^{U_{A}^{\beta_2}},
	(\bar{\psi}^{U_{\PP}^{\beta_3}})^{U_{A}^{\beta_2}},A)$. 
	Moreover $U(1)_A$ is a symmetry of $S_I$, thus $S_I (\psi^{\sutr},\bar{\psi}^{\sutr},A)  = 
	 S_I (\psi(x)^{U_{\PP}^{\beta_3}},\bar{\psi}(x)^{U_{\PP}^{\beta_3}},A)$.  Finally re-using the $U(1)_{\PP}$ invariance, we have  
	 $S_I (\psi^{\sutr},\bar{\psi}^{\sutr},A) = S_I (\psi,\bar{\psi},A)$. 
	 This shows the invariance of $S_I(\psi,\bar{\psi},A) $ under $\sun$ transformations with the restriction of gauge fields satisfying (\ref{eq:gauge}).

	\section{Chiralspin and Hamiltonian}\label{sec:ham}
	
	Another study, which we want to add, is the invariance of the free fermion Hamiltonian with respect $U(1)_{\PP}$ and $\sun$ (for the massless case) 
	and derive the $U(1)_{\PP}$ and $\sun$ transformations for creation and annihilation operators for fermions and anti-fermions. 
	Once we do this, we will briefly discuss the case where a gauge interaction is switched on. 
	\\
	
	Before to start, we point out here, that in this whole section \ref{sec:ham}, we assume that the spinor field $\psi$ describing free (and eventually massless $m=0$) fermions (or anti-fermions) is solution of the Dirac equation in the free case, i.e. $(\I\g{\mu}\partial_{\mu} - m)\psi(x)=0$, which is \cite{Mandl:1985bg} 
	
	\begin{equation}
	\begin{split}
	\psi(x)  
	= \sum_{r=0}^1 \int \frac{d^3 p}{(2\pi)^{3/2}} \left[c_r (\bm{p}) u_r (\bm{p})e^{-\I px} +d_r (\bm{p})^{\dagger} v_r (\bm{p})e^{\I px}  \right],
	\end{split}
	\label{eq:psi0}
	\end{equation}
	
	\noindent
	where $u_r(\bm{p})$ and $v_r(\bm{p})$ are reported in Eq. (\ref{eq:uv}), 
	and $c_r(\bm{p})$, $d_r(\bm{p})^{\dagger}$ are the annihilation and creation operators for particles and antiparticles respectively. 
	
	This is the particular situation where the free fermion action $S_F (\psi,\bar{\psi})$, calculated on such spinor field (\ref{eq:psi0}), reaches its minimum value, which is zero. 
	From such spinor field (\ref{eq:psi0}) we attempt to apply $U(1)_{\PP}$ and $\sun$ transformations, defined in the previous section in case of a totally generic spinor, in order to check the invariance of the free fermion Hamiltonian.
	
	This Hamiltonian, calculated using the spinor field in (\ref{eq:psi0}), is given by \cite{Mandl:1985bg} 
	
	\begin{equation}
		H_0 = \sum_{r=0}^1 \int d^3 p\, E_{\bm{p}} \left[c_r(\bm{p})^{\dagger}c_r(\bm{p}) + d_r(\bm{p})^{\dagger} d_r(\bm{p})\right],
		\label{eq:h}
	\end{equation}
	
	\noindent
	where $ E_{\bm{p}} = \sqrt{\vert\bm{p}\vert^2 + m^2}$. $H_0$ is invariant under parity transformation, 
	i.e. $PH_0 P^{\dagger} = H_0$. 
	This means that calling $c_r^P(\bm{p}) = P c_r (\bm{p})P^{\dagger}$ and $d_r^P(\bm{p})^{\dagger} = P d_r(\bm{p})^{\dagger} P^{\dagger}$, we have that 
	
	\begin{equation}
	\begin{split}
		H_0 &= \frac{1}{2}H_0 + \frac{1}{2}PH_0 P^{\dagger}\\
		 &= 
		\frac{1}{2}\sum_{r=0}^{1} \int d^3 p\,E_{\bm{p}}\left[
		c_r(\bm{p})^{\dagger}c_r(\bm{p}) + d_r(\bm{p})^{\dagger} d_r(\bm{p})\right]\\
		&+\frac{1}{2}\sum_{r=0}^{1} \int d^3 p\,E_{\bm{p}} \left[
		c_r^P(\bm{p})^{\dagger}c_r^P(\bm{p}) + d_r^P(\bm{p})^{\dagger} d_r^P(\bm{p})\right]\\
		&= \frac{1}{2}\sum_{r=0}^{1} \int d^3 p\,E_{\bm{p}}\left[
		C_r (\bm{p})^{\dagger}C_r (\bm{p}) + D_r (\bm{p})^{\dagger}D_r (\bm{p})\right],
	\end{split}
		\label{eq:h0}
	\end{equation}
	
	\noindent
	where we defined the following ``parity partners'' operators:
	
	\begin{equation}
		\begin{split}
		C_r (\bm{p}) = \left(\begin{matrix}
		c_r (\bm{p}) \\ c_r^P (\bm{p})
		\end{matrix}\right),\quad
		D_r (\bm{p})^{\dagger} = \left(\begin{matrix}
		d_r (\bm{p})^{\dagger} & d_r^P (\bm{p})^{\dagger}
		\end{matrix}\right).
		\end{split}
		\label{eq:cd}
	\end{equation}
	
	\noindent
	The expressions for $c_r (\bm{p})$ and $d_r (\bm{p})^{\dagger}$ can be obtained by the Fourier transform of $\psi(x)$ and they are reported in (\ref{eq:cdprop1}), while $c_r^P (\bm{p})$ and $d_r^P (\bm{p})^{\dagger}$ are obtained from the fact that $P\psi(x) P^{\dagger} = \g{0}\psi (\mathcal{P}x)$ and given in (\ref{eq:cdprop2}).
	
	\subsection{$U(1)_{\PP}$ and Hamiltonian}
	
	In order to check if $H_0$ is $U(1)_{\PP}$-invariant, we need to find how $c_r (\bm{p})$, $d_r (\bm{p})^{\dagger}$, $c_r^P (\bm{p})$ and $d_r^P (\bm{p})^{\dagger}$ transform. 
	For this purpose, we just need to use (\ref{eq:cdprop1}) and (\ref{eq:cdprop2}) together with the fact that $u_r(\bm{p})$ and $v_r(\bm{p})$ given in (\ref{eq:uv}) transform under parity as 
	
		\begin{equation}
	\g{0}u_r(\bm{p}) = u_r (-\bm{p}), \qquad
	\g{0}v_r(\bm{p}) = -v_r (-\bm{p}). 
	\label{eq:uvprop1}
	\end{equation}
	
	Here, we give the results
	
	\begin{equation}
		\begin{split}
		c_r (\bm{p})^{U_{\PP}^{\alpha}} &= \int \frac{d^3x}{(2\pi)^{3/2}} u_r(\bm{p})^{\dagger}\psi(x)^{U_{\PP}^{\alpha}}  e^{\I px}\\
		& =
		\cos(\alpha)\,c_r (\bm{p}) + \I\sin(\alpha)\,c_r^P (\bm{p}),\\
		c_r^P (\bm{p})^{U_{\PP}^{\alpha}} &= \int \frac{d^3x}{(2\pi)^{3/2}} u_r(-\bm{p})^{\dagger}\psi(x)^{U_{\PP}^{\alpha}}  e^{\I p(\mathcal{P}x)}\\
		& =
		\cos(\alpha)\,c_r^P (\bm{p}) + \I\sin(\alpha)\,c_r (\bm{p}),\\
		(d_r (\bm{p})^{\dagger})^{U_{\PP}^{\alpha}} &= \int \frac{d^3x}{(2\pi)^{3/2}} v_r(\bm{p})^{\dagger}\psi(x)^{U_{\PP}^{\alpha}}  e^{-\I px}\\
		& =
		\cos(\alpha)\,d_r (\bm{p})^{\dagger} + \I\sin(\alpha)\,d_r^P (\bm{p})^{\dagger},\\
		(d_r^P (\bm{p})^{\dagger})^{U_{\PP}^{\alpha}} &= \int \frac{d^3x}{(2\pi)^{3/2}} (-v_r(-\bm{p}))^{\dagger}\psi(x)^{U_{\PP}^{\alpha}}  e^{-\I p(\mathcal{P}x)}\\
		& =
		\cos(\alpha)\,d_r^P (\bm{p})^{\dagger} + \I\sin(\alpha)\,d_r (\bm{p})^{\dagger},\\
		\end{split}
		\label{eq:cpup}
	\end{equation}

	\noindent
	where we have just plugged the expression of $\psi(x)^{U_{\PP}^{\alpha}}$ of Eq. (\ref{eq:uid}) and expanded it. 
	For details on the calculations of (\ref{eq:cpup}), we refer to Appendix \ref{app:cpup}. 
	The result of Eq. (\ref{eq:cpup}) can be rewritten using the definition in (\ref{eq:cd}) as 
	
	\begin{equation}
		\begin{split}
		&C_r (\bm{p})\to C_r (\bm{p})^{U_{\PP}^{\alpha}} \equiv 
		\left(\begin{matrix}
		c_r (\bm{p})^{U_{\PP}^{\alpha}} \\
		c_r^P (\bm{p})^{U_{\PP}^{\alpha}}
		\end{matrix}\right) = e^{\I\alpha\sigma^1}C_r (\bm{p}),\\
				&D_r (\bm{p})^{\dagger}\to (D_r (\bm{p})^{\dagger})^{U_{\PP}^{\alpha}} \equiv 
		\left(\begin{matrix}
		(d_r (\bm{p})^{\dagger})^{U_{\PP}^{\alpha}}  & 
		(d_r^P (\bm{p})^{\dagger})^{U_{\PP}^{\alpha}}
		\end{matrix}\right)\\
		& \hspace{3.4cm}=  D_r (\bm{p})^{\dagger} e^{\I\alpha\sigma^1}\\
		\end{split}
		\label{eq:cpup1}
	\end{equation}
	
	\noindent
	where $\sigma^1$ is the 1st Pauli matrix acting on the 2-dimensional space defined in (\ref{eq:cd}). 
	As it is clear, the $H_0$ in (\ref{eq:h0}) is invariant under $U(1)_{\PP}$, since taking the hermitian of (\ref{eq:cpup1}), we obtain 
	$(C_r (\bm{p})^{U_{\PP}^{\alpha}})^{\dagger} = C_r (\bm{p})^{\dagger}\exp(-\I\alpha\sigma^1)$ 
	and $D_r (\bm{p})^{U_{\PP}^{\alpha}} \equiv ((D_r(\bm{p})^{\dagger})^{U_{\PP}^{\alpha}})^{\dagger} = \exp(-\I\alpha\sigma^1) D_r (\bm{p})$, 
	therefore $(C_r (\bm{p})^{U_{\PP}^{\alpha}})^{\dagger}C_r (\bm{p})^{U_{\PP}^{\alpha}} = C_r(\bm{p})^{\dagger}C_r (\bm{p})$ 
	and $(D_r (\bm{p})^{\dagger})^{U_{\PP}^{\alpha}}D_r (\bm{p})^{U_{\PP}^{\alpha}}=D_r(\bm{p})^{\dagger}D_r (\bm{p}) $.

	This result is actually pretty obvious in the free case, if you consider that $H_0$ commutes with $P$. 
	Take, for example, a state of a particle $\vert \bm{p}\rangle =  c_r(\bm{p})^{\dagger}\vert 0\rangle$ that has energy $E_{\bm{p}} = \langle \bm{p}\vert H_0 \vert \bm{p}\rangle$.
	The $U(1)_{\PP}$ invariance of $H_0$ tells us that the state $\vert \bm{p}\rangle$ is energetically equivalent to the state 
	$\vert \tilde{\bm{p}}\rangle = (c_r(\bm{p})^{U_{\PP}^{\alpha}})^{\dagger}\vert 0\rangle = (\cos(\alpha)c_r(\bm{p})^{\dagger} - \I\sin(\alpha)c_r^P(\bm{p})^{\dagger})\vert 0 \rangle$, where we used the second of Eq. (\ref{eq:cpup}). 
	We rewrite it as $\vert \tilde{\bm{p}}\rangle = \cos(\alpha)\vert \bm{p}\rangle - \I\sin(\alpha)\vert -\bm{p}\rangle$, because $c_r^P(\bm{p})^{\dagger}\vert 0 \rangle = Pc_r(\bm{p})^{\dagger}\vert 0 \rangle = P\vert \bm{p} \rangle = \vert -\bm{p}\rangle$. 
	Hence we have $\langle \tilde{\bm{p}}\vert H_0\vert \tilde{\bm{p}}\rangle = \cos(\alpha)^2 \langle \bm{p}\vert H_0\vert \bm{p}\rangle + \I\sin(\alpha)\cos(\alpha)\langle -\bm{p}\vert H_0\vert \bm{p}\rangle 
	-\I\sin(\alpha)\cos(\alpha)\langle \bm{p}\vert H_0\vert -\bm{p}\rangle
	+\sin(\alpha)^2\langle -\bm{p}\vert H_0\vert -\bm{p}\rangle$. 
	However $\langle -\bm{p}\vert H_0\vert -\bm{p}\rangle = \langle \bm{p}\vert P H_0 P^{\dagger}\vert \bm{p}\rangle = \langle \bm{p}\vert H_0\vert \bm{p}\rangle = E_{\bm{p}}$, because $[H_0,P] = 0$, and for the same reason $\langle -\bm{p}\vert H_0\vert \bm{p}\rangle  = \langle \bm{p}\vert H_0\vert - \bm{p}\rangle $. 
	Therefore $\langle \tilde{\bm{p}}\vert H_0\vert \tilde{\bm{p}}\rangle = \langle \bm{p}\vert H_0 \vert \bm{p}\rangle = E_{\bm{p}}$, for whatever value of $\bm{p}$. 
	Hence the two states have the same energy.
	Basically in the free case, the commutation of between Hamiltonian and parity gives place to the $U(1)_{\PP}$ invariance.
	
	
	\subsection{$\sun$ and Hamiltonian}\label{sec:sunham}
	
	We prove now that the Hamiltonian $H_0$ is also invariant under $\sun$ transformations for $m=0$. 
	In order to see this point, we give how $c_r (\bm{p})$, $c_r^P (\bm{p})$, $d_r (\bm{p})^{\dagger}$, $d_r^P (\bm{p})^{\dagger}$ transform. 
	At first we define the fields $U_r (\bm{p})^{\dagger}  = \left(\begin{matrix}
	u_r (\bm{p})^{\dagger}  & 0
	\end{matrix}\right)$ and 
	$V_r (\bm{p})^{\dagger} = \left(\begin{matrix}
	v_r (\bm{p})^{\dagger}  & 0
	\end{matrix}\right)$, 
	which are defined in the ``parity partners'' space, as $\Psi(x)$ in (\ref{eq:psi}).
	Hence we have that $u_r(\bm{p})^{\dagger}\psi(x) = U_r (\bm{p})^{\dagger}\spiu\Psi(x)$ 
	and 
	$v_r(\bm{p})^{\dagger}\psi(x) = V_r (\bm{p})^{\dagger}\spiu\Psi(x)$, 
	where $\mathds{1}$ acts in the Dirac space, $\Psi(x)$ is given in (\ref{eq:psi}) and $(\mathds{1}+\sigma^1)$ acts on the 2-dimensional space of ``parity partners''. 
	Secondly, we use that for $m=0$, the vectors $u_r$ and $v_r$ satisfy the properties: $\g{5} u_r (\bm{p}) = (\bm{\sigma}\cdot\bm{p}/\vert \bm{p} \vert)\,u_r(\bm{p})$ and $\g{5} v_r (\bm{p}) = (\bm{\sigma}\cdot\bm{p}/\vert \bm{p} \vert)\,v_r(\bm{p})$. 
	This reflects the fact that in the massless case $\g{5}$ coincides with the helicity operator $\bm{\sigma}\cdot\bm{p}/\vert \bm{p} \vert$. 
	For convenience we choose $\chi_r$ and consequently $\chi'_r$, defined by Eq. (\ref{eq:uv}) in the solution of the Dirac equation, such that they are eigenstates of the helicity operator, i.e. 
	$(\bm{\sigma}\cdot\bm{p}/\vert \bm{p} \vert)\,\chi_0 = \chi_0$ and 
	$(\bm{\sigma}\cdot\bm{p}/\vert \bm{p} \vert)\,\chi_1 = -\chi_1$. 
	This means that for $m=0$ we have 
	
	\begin{equation}
	\g{5}u_r(\bm{p}) = h_{r}\,u_r (\bm{p}), \qquad
	\g{5}v_r(\bm{p}) = h_{r+ 1}\,v_r (\bm{p}), 
	\label{eq:uvprop2}
	\end{equation}
	
	\noindent
	with $h_r = (-1)^r$, helicity of the particle. 
	
	Using these two considerations, we can redefine $c_r (\bm{p})$, $c_r^P (\bm{p})$, $d_r (\bm{p})^{\dagger}$ and $d_r^P (\bm{p})^{\dagger}$, from Eqs. (\ref{eq:cdprop1}) and (\ref{eq:cdprop2}), in terms of $\Psi$, $V_r$ and  $U_r$ and in order to look their $\sun$ transformation we use that from Eq. (\ref{eq:su2csnew}), we can write $\sutr$ extensively as 
	
	\begin{equation}
		\sutr = \cos(\alpha) + \I\sin(\alpha)\left[e_1 \Sigma_1^{\mathcal{P}}+e_2 \Sigma_2^{\mathcal{P}}+e_3 \Sigma_3^{\mathcal{P}}\right],
		\label{eq:sutr}
	\end{equation}
	
	\noindent
	where $(\alpha_1,\alpha_2,\alpha_3) = \alpha(e_1,e_2,e_3)$, with $\sum_{i=1}^3 e_i^2 = 1$.
	
	Therefore for $c_r (\bm{p})$ and $c_r^P (\bm{p})$ we get
	
	\begin{widetext}
	\begin{equation}
		\begin{split}
		&c_r(\bm{p})^{\sutr} = \int \frac{d^3 x}{(2\pi)^{3/2}}U_r (\bm{p})^{\dagger}\spiu\Psi(x)^{\sutr} e^{\I px}\\
		&=\cos(\alpha)\,c_r(\bm{p}) + \I \sin(\alpha)\,\left[e_1 c_r^P(\bm{p}) + e_2 \I h_r c_r^P(\bm{p}) - e_3 h_r c_r(\bm{p})\right],\\
		&c_r^P(\bm{p})^{\sutr} = \int \frac{d^3 x}{(2\pi)^{3/2}}U_r (-\bm{p})^{\dagger}\spiu\Psi(x)^{\sutr} e^{\I p(\mathcal{P}x)}\\
		&=\cos(\alpha)\,c_r^P(\bm{p}) + \I \sin(\alpha)\,\left[e_1 c_r(\bm{p}) - e_2\I h_r c_r(\bm{p}) + e_3 h_r c_r^P(\bm{p})\right],
		\end{split}
		\label{eq:csu2cs}
	\end{equation}
	\end{widetext}
	
	\noindent
	in which we used the change of variables $x\to\mathcal{P}x$, whenever necessary, and that by definition $\Psi(\mathcal{P}x) = (\sigma^3\otimes\mathds{1}) \Psi(x)$. 
	Details regarding the derivation of (\ref{eq:csu2cs}) are given in Appendix \ref{app:cdsu2cs}.
	
	Eq. (\ref{eq:csu2cs}) can be written in a compact way using the notation in (\ref{eq:cd}) as 
	
		\begin{equation}
	\begin{split}
	&C_r (\bm{p})\to C_r (\bm{p})^{\sutr} \equiv 
	\left(\begin{matrix}
	c_r (\bm{p})^{\sutr} \\
	c_r^P (\bm{p})^{\sutr}
	\end{matrix}\right) = e^{\I\alpha_n\Sigma_{n(c)}^{\mathcal{P}}}C_r (\bm{p}),\\
	\end{split}
	\label{eq:csu2cs2}
	\end{equation}
	
	\noindent
	where $\Sigma_{n(c)}^{\mathcal{P}} = \{\sigma^1,-\sigma^2 h_r,-\sigma^3 h_r\}$, are all traceless, hermitian and with the property:  $[\Sigma_{i(c)}^{\mathcal{P}},\Sigma_{j(c)}^{\mathcal{P}}] = 2\I \epsilon_{ijk}\Sigma_{k(c)}^{\mathcal{P}}$, since $h_r^2 = 1$. 
	Therefore we have found a representation of $\sun$ for the  transformations of $C_r (\bm{p})$ in the massless case. Notice that (\ref{eq:csu2cs2}) are basically rotations in the space of the ``parity partners'': $c_r(\bm{p})$ and $c_r^P(\bm{p})$, which takes into account the helicity of our particles. 
	
	The same can be done for $d_r (\bm{p})^{\dagger}$ and $d_r^P (\bm{p})^{\dagger}$ and we obtain the following result
	
	\begin{widetext}
	\begin{equation}
		\begin{split}
	&(	d_r (\bm{p})^{\dagger})^{\sutr} = \int \frac{d^3 x}{(2\pi)^{3/2}}V_r (\bm{p})^{\dagger}\spiu\Psi(x)^{\sutr}e^{-\I px}\\
	&=\cos(\alpha)\,d_r (\bm{p})^{\dagger} + \I\sin(\alpha)\left[e_1 d_r^P (\bm{p})^{\dagger} + e_2 \I h_{r+ 1}\,d_r^P (\bm{p})^{\dagger} - e_3 h_{r+ 1}d_r (\bm{p})^{\dagger}\right],\\
		&(	d_r^P (\bm{p})^{\dagger})^{\sutr} = \int \frac{d^3 x}{(2\pi)^{3/2}}(-V_r (-\bm{p}))^{\dagger}\spiu \Psi(x)^{\sutr}e^{-\I p(\mathcal{P}x)}\\
	&=\cos(\alpha)\,d_r^P (\bm{p})^{\dagger} + \I\sin(\alpha)\left[e_1 d_r (\bm{p})^{\dagger} - e_2 \I h_{r+ 1}\,d_r (\bm{p})^{\dagger} + e_3  h_{r+ 1}d_r^P (\bm{p})^{\dagger}\right],\\
		\end{split}
		\label{eq:dsu2cs}
	\end{equation}
	\end{widetext}
	
	\noindent
	where we used the same procedure as before. The details of this calculations are again in Appendix \ref{app:cdsu2cs}. 
	Eq. (\ref{eq:dsu2cs}) can be given in a compact way as 
	
	\begin{equation}
	\begin{split}
			D_r (\bm{p})^{\dagger}\to (D_r (\bm{p})^{\dagger})^{\sutr} &\equiv 
		\left(\begin{matrix}
		(d_r (\bm{p})^{\dagger})^{\sutr}  
		(d_r^P (\bm{p})^{\dagger})^{\sutr}
		\end{matrix}\right)\\
		& = D_r (\bm{p})^{\dagger} \left(e^{\I\alpha_n\Sigma_{n(d)}^{\mathcal{P}}}\right)^T,
		\end{split}
	\label{eq:dsu2cs2}
	\end{equation}
	
	\noindent
	where $\Sigma_{n(d)}^{\mathcal{P}} = \{\sigma^1,-\sigma^2 h_{r+ 1},-\sigma^3 h_{r+ 1}\}$, are again all traceless, hermitian and with the property:  $[\Sigma_{i(d)}^{\mathcal{P}},\Sigma_{j(d)}^{\mathcal{P}}] = 2\I \epsilon_{ijk}\Sigma_{k(d)}^{\mathcal{P}}$, since $h_{r+ 1}^2 = 1$. 
	Eq. (\ref{eq:dsu2cs2}) expresses the representation of $\sun$ transformations for $D_r(\bm{p})$ and it is a rotation of the ``parity partners'': $d_r(\bm{p})^{\dagger}$ and $d_r^P(\bm{p})^{\dagger}$, where we take into account the helicity for antiparticles. 
	
	Moreover, we can obtain the $\sun$-transformations of $C_r (\bm{p})^{\dagger}$ and $D_r (\bm{p})$ considering $(C_r (\bm{p})^{\dagger})^{\sutr} = (C_r (\bm{p})^{\sutr})^{\dagger}$ and $D_r (\bm{p})^{\sutr} = ((D_r (\bm{p})^{\dagger})^{\sutr})^{\dagger}$.
	
	As we can observe, $H_0$ in (\ref{eq:h0}) is of course invariant under  transformations of $C_r (\bm{p})$ and $D_r (\bm{p})^{\dagger}$ given in (\ref{eq:csu2cs2}) and (\ref{eq:dsu2cs2}), 
	because $(C_r (\bm{p})^{\sutr})^{\dagger}C_r (\bm{p})^{\sutr} = C_r (\bm{p})^{\dagger}C_r (\bm{p})$ and $(D_r (\bm{p})^{\dagger})^{\sutr}D_r(\bm{p})^{\sutr} = D_r (\bm{p})^{\dagger}D_r (\bm{p})$. 
	This conclude our proof that $H_0$ at $m=0$ is $\sun$-invariant. 
	
	It remains to see what happens when $m\neq 0 $ and in presence of a gauge interaction. 
	For $m\neq 0 $, we already know that $H_0$ is not invariant under $U(1)_A$ and therefore is not invariant under $\sun$, since $U(1)_A \subset\sun$. 
	The case of a gauge interaction just makes fall the relation (\ref{eq:uvprop1}), valid in the free case. 
	This means that the relations (\ref{eq:csu2cs2}) and (\ref{eq:dsu2cs2}) do not represent anymore $\sun$ transformations.
	Hence in this situation we do not expect to have this symmetry, even because, as we have already seen in the previous section, the gauge interaction breaks the invariance also in the action, because $U(1)_{\PP}$ is broken explicitly. 
	
	\section{Mass degeneration and symmetry}\label{sec:correlators} 
	
	Now we consider the case of QCD, where the fermion fields $\psi$ that we have discussed so far, are interpreted as quark fields.
	 
	The $\su$ group transformations, as has been described in \cite{Denissenya:2014poa,Denissenya:2014ywa,Denissenya:2015mqa,Glozman:2017dfd,Glozman:2016ayd}, can be defined in Minkowskian space-time as 
	
	\begin{equation}
	\begin{split}
	\su\,:\,&\psi(x) \TO \psi(x)^{\suno} = \suno\,\psi(x),\\
	& \suno = e^{\I\alpha_n\Sigma_n}\in\su,
	\end{split}
	\label{eq:oldsu2cs}
	\end{equation}
	
	\noindent
	where $\Sigma_n = \{\g{0},\I\g{5}\g{0},-\g{5}\}$, and we have just substituted $\g{4}\TO\g{0}$ from Euclidean to Minkowskian space-time. 
	We can see from the above definition that we have two important subgroups just tuning the $\alpha_n$s. 
	One is $U(1)_A$ (which is also a subgroup of $\sun$) and the other is the $U(1)$ group generated by $\g{0}$, from which the group transformation is obtained by choosing $(\alpha_1,\alpha_2,\alpha_3) = (\alpha,0,0)$ in (\ref{eq:oldsu2cs}). Hence we get 
	
	\begin{equation}
	\begin{split}
		U(1)_{0}\,:\,&\psi(x)\to\psi(x)^{U(1)_{0}^{\alpha}}\\
		& = e^{\I\alpha\g{0}}\psi(x) = \cos(\alpha)\psi(x) + \I\sin(\alpha)\g{0}\psi(x).
	\end{split}
		\label{eq:u10}
	\end{equation}	
	
	\noindent
	Now, using the Euler decomposition, whatever element of $\su$ can be obtained by the product of three matrices belonging to $U(1)_A$ and $U(1)_{0}$. 
	Therefore the real difference between $\su$ and $\sun$ lies in their different subgroups: $U(1)_{0}$, given in (\ref{eq:u10}), and $U(1)_{\PP}$, given in (\ref{eq:uid}), respectively. 
	However while $U(1)_{\PP}$ is a symmetry of the free fermion action $S_F$ in (\ref{eq:sf}), $U(1)_0$ is broken explicitly. 
	This is why $\su$ is not a symmetry of free massless quarks.
	Now we want to show how $	\sun$ is related to $\su$ and its consequence on hadron correlators and mass degeneration. 
	At first we take the case when $\Psi(x)$ is evaluated at the point $x^{(t)} = (x_0,\bm{0})$. 
	In this situation, by definition $\psi_{+}(x^{(t)}) = \psi(x^{(t)})$ and $\psi_{-}(x^{(t)}) = 0$. 
	Therefore the transformation (\ref{eq:su2csnew}) becomes 
	
	\begin{equation}
	\begin{split}
	\sun\,:
	&\Psi(x^{(t)})\to\Psi(x^{(t)})^{\sutr}\\ 
	&= e^{\I\alpha_n\Sigma_n^{\mathcal{P}}}
	\left(\begin{matrix}
	\psi(x^{(t)})\\ 0
	\end{matrix}\right)= 
	\left(\begin{matrix}
	e^{\I\alpha_n\Sigma_n}\psi(x^{(t)})\\ 0
	\end{matrix}\right).
	\end{split}
	\label{eq:sucs2_xt}
	\end{equation} 
	
	\noindent
	Thus $\psi_{-}(x^{(t)})^{\sutr} = 0$ and $\psi_{+}(x^{(t)})^{\sutr} = \exp(\I\alpha_n\Sigma_n)\psi(x^{(t)})$, 
	where the $\Sigma_n$s are the $\su$ generators previously introduced. 
	
	Now naming for simplicity $\psi(x^{(t)})^{\sutr} \equiv \psi_{+}(x^{(t)})^{\sutr}$, 
	the \eq (\ref{eq:sucs2_xt}) can be rewritten also as 
	
	\begin{equation}
	\sun\,:\,\psi(x^{(t)}) \TO \psi(x^{(t)})^{\sutr} = e^{\I\alpha_n\Sigma_n}\psi(x^{(t)}).
	\label{eq:sucs2_xt_2}
	\end{equation}
	
	\noindent
	It coincides with \eq (\ref{eq:oldsu2cs}) which means that $\psi(x^{(t)})^{\suno}=\psi(x^{(t)})^{\sutr}$. 
	Hence $\sun$ and $\su$ are indistinguishable when they act on the spinor $\psi(x^{(t)})$. 
	This has an important consequence as we start to describe now. 
	Take an hadron observable $O_H(x)$ made by $N_q$ quarks and $\bar{N}_q$ antiquarks, i.e.  
	
	\begin{equation}
		O_H(x) = \mathcal{H}_{i_1,..,i_{N_q},j_1,...,j_{\bar{N}_q}} \prod_{l=1}^{N_q} \psi(x)_{i_l}\prod_{k=1}^{\bar{N}_q} \bar{\psi}_{j_k}(x)
		\label{eq:oh}
	\end{equation}

	\noindent
	(for example for $N_q=3$ and $\bar{N}_q=0$ we get a baryon and $N_q=\bar{N}_q=1$ we get a meson), 
	where $\mathcal{H}$ is a tensor specifying the quantum numbers of the hadron $H$, and the indices $\{i\}_{l=1,...N_q}$ and $\{j\}_{k=1,...\bar{N}_q}$ enclose Dirac, flavor and eventually  color indices.
	We now choose to transform it with $\su$ and for some choice of the parameters $\alpha_n$s in (\ref{eq:oldsu2cs}) we get another hadron observable, i.e. 
	
	\begin{equation}
		O_{H'}(x) = \mathcal{H'}_{i_1,..,i_{N_q},j_1,...,j_{\bar{N}_q}} \prod_{l=1}^{N_q} \psi(x)_{i_l}\prod_{k=1}^{\bar{N}_q} \bar{\psi}_{j_k}(x),
		\label{eq:ohp}
	\end{equation}

	\noindent
	which is the observable for the hadron $H'$. 
	$O_H (x)$ and $O_{H'}(x)$ in (\ref{eq:oh}) and (\ref{eq:ohp}) are connected via $\su$ if for some $\bm{\alpha}$ we have   
	$O_{H'}(x) = O_{H}(x)^{\suno} \equiv \mathcal{H}_{i_1,..,i_{N_q},j_1,...,j_{\bar{N}_q}} \prod_{l=1}^{N_q} \psi(x)_{i_l}^{\suno}\prod_{k=1}^{\bar{N}_q} \bar{\psi}_{j_k}(x)^{\suno}$.  
	Now, since for $x=x^{(t)}$ we have $\psi(x^{(t)})^{\suno}=\psi(x^{(t)})^{\sutr}$, 
	then $O_{H}(x)^{\sutr}$, which is the $\sun$ transformation of $O_H (x)$ with the same set of parameters $\alpha_n$s has the property $O_{H}(x^{(t)})^{\suno}=O_{H}(x^{(t)})^{\sutr}$ 
	and consequently $O_{H'}(x^{(t)}) =O_{H}(x^{(t)})^{\sutr} $. 
	Therefore even if $O_{H}(x)^{\sutr}$ could be not associated with an hadron for a generic $x$, 
	however at $x=x^{(t)}$, it coincides with the hadron operator $O_{H'}(x^{(t)})$. 
	At this point, suppose to consider these two hadron correlators and their expansion in the energy eigenstates using the translational invariance $O_H (x^{(t)}) = \exp(-\I H x_0)O_H (0)\exp(\I H x_0)$, we have
	
	\begin{equation}
	\begin{split}
	& \langle 0 \vert O_H (y^{(t)}) O_H (x^{(t)})^{\dagger}\vert 0 \rangle = \sum_n \vert\langle 0 \vert O_H (0)\vert n\rangle\vert^2 e^{-\I E_n \mathcal{T}},\\
	&\langle 0 \vert O_{H'} (y^{(t)}) O_{H'} (x^{(t)})^{\dagger}\vert 0 \rangle = \sum_n \vert\langle 0 \vert O_{H'} (0)\vert n\rangle\vert^2 e^{-\I E'_n \mathcal{T}},
	\end{split}
	\label{eq:corr}
	\end{equation}
	
	\noindent
	where $y^{(t)}  = (y_0,\bm{0})$, $\mathcal{T} =  y_0 - x_0$, while $m_H = E_0$ and $m_{H'} = E'_0$ are the masses associated with the hadrons $H$ and $H'$ respectively. 
	Now, if $\su$ is a symmetry of the theory (which seems to be in truncated studies \cite{Denissenya:2014poa,Denissenya:2014ywa,Denissenya:2015mqa}
	then $\langle 0 \vert O_H (y^{(t)})^{\suno} (O_H (x^{(t)})^{\suno})^{\dagger}\vert 0 \rangle = \langle 0 \vert O_H (y^{(t)}) O_H (x^{(t)})^{\dagger}\vert 0 \rangle$ 
	and therefore since we have chosen the $\su$ transformations such that 
	$O_H(x)^{\suno} = O_{H'}(x)$, then we have $m_{H'} = m_{H}$, from (\ref{eq:corr}).  
	This means that a degeneration of masses appears. 
	Let us see the opposite, i.e. we find a degeneration $m_{H'} = m_{H}$ coming from 
	$\langle 0 \vert O_{H'} (y^{(t)})O_{H'} (x^{(t)})^{\dagger}\vert 0 \rangle = \langle 0 \vert O_H (y^{(t)}) O_H (x^{(t)})^{\dagger}\vert 0 \rangle$.
	In that case at $x = x^{(t)}$ (and also $y = y^{(t)}$), we have $O_H(x^{(t)})^{\sutr} = O_H (x^{(t)})^{\suno}$, which means from the correlator side that 
	
	\begin{equation}
	\begin{split}
	&\langle 0 \vert O_{H'} (y^{(t)}) O_{H'} (x^{(t)})^{\dagger}\vert 0 \rangle\\
	=&\langle 0 \vert O_H (y^{(t)})^{\suno} (O_H (x^{(t)})^{\suno})^{\dagger}\vert 0 \rangle\\ 
	=&\langle 0 \vert O_H (y^{(t)})^{\sutr} (O_H (x^{(t)})^{\sutr})^{\dagger}\vert 0 \rangle \\
	=& \langle 0 \vert O_H (y^{(t)}) O_H (x^{(t)})^{\dagger}\vert 0 \rangle.
	\end{split}
	\end{equation}

	\noindent
	This implies that $\sun$ symmetry can also explain the same mass degeneration. 
	Therefore looking just the mass degeneration $m_{H'} = m_H$ does not tell us if the symmetry is $\su$ of the truncated studies \cite{Denissenya:2014poa,Denissenya:2014ywa,Denissenya:2015mqa,Rohrhofer:2017grg,Rohrhofer:2019qwq,Rohrhofer:2019qal,Glozman:2021jlk,Glozman:2017dfd,Glozman:2016ayd}, where at first the mass degeneration has been observed, or $\sun$ of this paper. 
	
	However because $\sun$ is a symmetry of free quarks (in the massless case), its possible presence does not go in contrast with the deconfinement regime at high temperature QCD, 
	since in section \ref{sec:chiralspin2} we have shown that it is a symmetry of the action as well the Hamiltonian of free fermions (let say quarks) in the massless case.
	On the contrary 
	$\su$ is not a symmetry of $S_F$ at $m=0$ because $U(1)_{0}$ is explicitly broken (see Refs. \cite{Glozman:2017dfd,Glozman:2016ayd}). 
	Therefore we expect that $\sun$ to be visible at temperature $T\to\infty$, where quarks will behave as quasi free particles.
	Hence, while $\su$ is just present in the range $T_c < T \lesssim 3 T_c $ \cite{Rohrhofer:2017grg,Rohrhofer:2019qwq,Rohrhofer:2019qal,Glozman:2021jlk}, $\sun$ could be visible (at least approximately) also at $T> 3T_c$ and describe the large mass degeneracy which was previously explained by $\su$ in the  truncated studies \cite{Denissenya:2014poa,Denissenya:2014ywa,Denissenya:2015mqa,Glozman:2017dfd,Glozman:2016ayd}. 
	Therefore a study on the lattice on this point is strongly suggested. 
	
	Now, lattice calculations, as we know, are generally performed at zero chemical potential (due to technical difficulties). 
	However, if we suppose to switch on an eventual chemical potential term in the action, 
	this would not spoil $\sun$. In fact the $\sun$ transformations leave also invariant the chemical potential part of the action (as also $\su$ does, see Refs. \cite{Glozman:2017dfd,Glozman:2019fku}), 
	which is $S_{(\mu)}(\psi,\bar{\psi}) = \mu\int d^4 x\,\psi(x)^{\dagger}\psi(x)$. 
	 	The demonstration starts rewriting it as 
	$S_{(\mu)}(\psi,\bar{\psi}) = \mu\int d^4 x\, (\psi_{+}(x)^{\dagger}\psi_{+}(x) +\psi_{-}(x)^{\dagger}\psi_{-}(x))$, since $\psi(x) = \psi_{+}(x) + \psi_{-}(x)$. 
	This because the mixing terms give zero under the integration, indeed 
	$\int d^4 x\,\psi_{\pm}(x)^{\dagger}\psi_{\mp}(x) = \int d^4 x\,\psi_{\pm}(\mathcal{P}x)^{\dagger}\psi_{\mp}(\mathcal{P}x) =  -\int d^4 x\,\psi_{\pm}(x)^{\dagger}\psi_{\mp}(x) = 0$, where firstly we used the change of variable $x\to\mathcal{P}x$ and $\vert\det(\mathcal{P})\vert=1$ 
	and secondly that $\psi_{\pm}(\mathcal{P}x)=\pm\psi_{\pm}(x)$ by definition. 
	Therefore, using (\ref{eq:psi}), then we have $S_{(\mu)}(\psi,\bar{\psi}) = \mu\int d^4 x\,\Psi(x)^{\dagger}\Psi(x)$. 
	Consequently from Eq. (\ref{eq:su2csnew}), we get  $S_{(\mu)}(\psi^{\sutr},\bar{\psi}^{\sutr}) = 
	\mu\INT (\Psi(x)^{\sutr})^{\dagger}\Psi(x)^{\sutr} = 
	\mu\INT\Psi(x)^{\dagger}\Psi(x)=S_{(\mu)}(\psi,\bar{\psi})$. Hence the  invariance of $S_{(\mu)}(\psi,\bar{\psi})$ under $\sun$ is proven. 
	Consequently, if there is a regime at high temperature where $\sun$ is a symmetry of QCD, then a possible non zero chemical potential do not break such symmetry.
	
	\section{Summary}\label{sec:summary}
	
	In Ref. \cite{Catillo:2021rrq}, we have seen how $\sun$ can be a candidate for describing the large degeneracy found on lattice calculations \cite{Denissenya:2014poa,Denissenya:2014ywa,Denissenya:2015mqa}, and eventually at high temperature QCD studies \cite{Rohrhofer:2017grg,Rohrhofer:2019qwq,Rohrhofer:2019qal,Glozman:2021jlk}. 
	Here, we have defined such group in Minkowskian and proved that the fermionic action of free massless fermions is left invariant under such group and a chemical potential term is also $\sun$-invariant. 
	
	In the case of a gauge interaction, $\sun$ is explicitly broken except for some special cases \cite{Catillo:2021rrq}. 
	Therefore a more profound investigation on this is needed. 
	We have also seen that the Hamiltonian of free fermions is $\sun$-invariant and given how the creation and annihilation operators of fermions and anti-fermions, when organized in ``parity partners'' (in the sense of Eq. (\ref{eq:cd})), would transform in this case, see Eqs. (\ref{eq:csu2cs2}) and (\ref{eq:dsu2cs2}).
	
	Moreover, from the Minkowskian perspective as we have done in Euclidean \cite{Catillo:2021rrq}, we have seen that a mass degeneration given by a possible presence of the \textit{chiralspin} group $\su$ symmetry can be either explained by our \textit{chiralspin} group $\sun$. 
	However since $\sun$ is a symmetry of free fermions, then it is compatible with the presence of the deconfinement regime in QCD. 
	Therefore it could be visible (at least \textit{effectively}) at high temperature QCD, i.e. $T> T_c$ (as e.g. $U(1)_A$ in Refs. \cite{Bazavov:2012qja,Cossu:2013uua,Tomiya:2016jwr}), but this is still something to be checked on lattice calculations. 
	
	As we have shown in this paper, 
	$\su$ and $\sun$ have both $U(1)_A$ as subgroup, nevertheless
	 they differ each other by the subgroup generated by $\g{0}$, 
	i.e. $U(1)_{0}\subset\su$, look Eq. (\ref{eq:u10}), and $U(1)_{\PP}\subset\sun$, look Eq. (\ref{eq:uid}). 
	Now, $U(1)_A$ has been already studied on the lattice and the suppression of its breaking for extremely high temperature is pretty evident by many lattice studies \cite{Bazavov:2012qja,Cossu:2013uua,Tomiya:2016jwr}. 
	Therefore in order to check the presence of $\sun$ in QCD at high $T$, is sufficient to verify the presence of $U(1)_{\PP}$ (defined by Eq. (\ref{eq:uid})) in lattice QCD at high temperature, 
	which still haven't been done yet.

	\appendix
	
		\section{Known formulas} \label{app:not}
	
	In this paper we have used the Dirac representation for the gamma matrices, namely 
	
	\begin{equation}
	\g{0} = \left(\begin{matrix}
	\mathds{1} & 0\\
	0 & -\mathds{1}
	\end{matrix}\right),\quad
	\g{k} = \left(\begin{matrix}
	0 & \tau^k\\
	-\tau^k & 0
	\end{matrix}\right),\quad
	\g{5} = \left(\begin{matrix}
	0 & \mathds{1}\\
	\mathds{1} & 0
	\end{matrix}\right),
	\label{eq:notgamma}
	\end{equation}
	
	\noindent
	where $\tau^k$ for $k=1,2,3$ are the Pauli matrices.
	Using such representation the solution of the Dirac equation: $(\I\g{\mu}\partial_{\mu}^x - m)\,\psi(x)=0$, given in (\ref{eq:psi0}), 
	contains the 4-component vectors $u_r (\bm{p})$ and $v_r (\bm{p})$, which have the following structure \cite{Mandl:1985bg}
	
	\begin{equation}
	\begin{split}
	&u_r (\bm{p}) = \sqrt{\frac{E+m}{2E}}\left(\begin{matrix}
	\chi_r\\ \frac{\bm{\sigma}\cdot\bm{p}}{E+m} \chi_r
	\end{matrix}\right),\\
	&v_r (\bm{p}) = \sqrt{\frac{E+m}{2E}}\left(\begin{matrix}
	\frac{\bm{\sigma}\cdot\bm{p}}{E+m} \chi'_r \\ 	\chi'_r
	\end{matrix}\right),
	\end{split}
	\label{eq:uv}
	\end{equation}
	
	\noindent
	with $\chi'_r = \chi_{r\oplus 1}$ (where $r\oplus 1 = (r + 1)\mod 2$), while $\chi_0$ and $\chi_1$ are two two-dimensional orthogonal vectors, i.e. $\chi_r^{\dagger}\chi_{r'}=\chi_r^{'\dagger}\chi'_{r'} = \delta_{r r'}$. 
	Consequently the normalization is $u_r (\bm{p})^{\dagger}u_{r'} (\bm{p})=v_r (\bm{p})^{\dagger}v_{r'} (\bm{p}) = \delta_{r r'}$, for whatever $\bm{p}$. 
	
	\noindent
	Other interesting features come from the coefficients $c_r (\bm{p})$ and $d_r (\bm{p})^{\dagger}$ in (\ref{eq:psi0}), namely the annihilation and creation operators of particles and antiparticles respectively. 
	They can be obtained by the Fourier transform 
	
	\begin{equation}
	\begin{split}
	&c_r (\bm{p}) = \int \frac{d^3 x}{(2\pi)^{3/2}} u_r(\bm{p})^{\dagger}\psi(x) \,e^{\I px},\\
	&d_r (\bm{p})^{\dagger} = \int \frac{d^3 x}{(2\pi)^{3/2}} v_r(\bm{p})^{\dagger}\psi(x) \,e^{-\I px}.
	\end{split}
	\label{eq:cdprop1}
	\end{equation}
	
	\noindent
	Moreover from the parity transformation $P\psi(x)P^{\dagger} = \g{0}\psi(\mathcal{P}x)$ and the relations (\ref{eq:uvprop1}), 
	we can express $c_r^P(\bm{p})\equiv P c_r (\bm{p})P^{\dagger}$ and 
	$d_r^P(\bm{p})^{\dagger}\equiv P d_r (\bm{p})^{\dagger}P^{\dagger}$ as 
	
	\begin{equation}
	\begin{split}
	&c_r^P (\bm{p}) = \int \frac{d^3 x}{(2\pi)^{3/2}} u_r(-\bm{p})^{\dagger}\psi(x) \,e^{\I p(\mathcal{P}x)},\\ 
	&d_r^P (\bm{p})^{\dagger} = \int \frac{d^3 x}{(2\pi)^{3/2}} (-v_r(-\bm{p}))^{\dagger}\psi(x) \,e^{-\I p(\mathcal{P}x)}.
	\end{split}	
	\label{eq:cdprop2}
	\end{equation}
	
	\noindent
	where we used that $\g{0} = (\g{0})^{\dagger}$, changed the variable $x \to \mathcal{P}x$ used that $\mathcal{P}= \diag(1,-1,-1,-1) = \mathcal{P}^{-1}$ and that the Jacobian $\vert\det(P)\vert^2 = 1$. 

	\section{$U(1)_{\PA}$ breaking}\label{app:b}
	
	In this Appendix we want to prove that the gauge interaction and also the mass term in the action break explicitly $U(1)_{\PA}$, subgroup of $\sun$. 
	The expression of $U(1)_{\PA}$ transformations for a fermion field $\psi$  is given in Eq. (\ref{eq:u1pa}), from which the transformation for $\bar{\psi}$ is $\bar{\psi}(x)^{U_{\PA}^{\alpha}} = (\psi(x)^{U_{\PA}^{\alpha}})^{\dagger}\g{0} = \cos(\alpha)\,\bar{\psi}(x) + \I\sin(\alpha)\,\bar{\psi}(\mathcal{P}x)\,\I\g{5}\g{0}$. 
	
	Let us start with the gauge interaction term of the action $S_I (\psi,\bar{\psi},A) = g \INT\, \bar{\psi}(x)\g{\mu}A_{\mu}(x)\psi(x)$, i.e.
		
	\begin{equation}
	\begin{split}
		&S_I (\psi^{U_{\PA}^{\alpha}},\bar{\psi}^{U_{\PA}^{\alpha}},A)/g\\
		 &= \INT\, \bar{\psi}(x)^{U_{\PA}^{\alpha}}\g{\mu}A_{\mu}(x)\psi(x)^{U_{\PA}^{\alpha}}\\
		&= \cos(\alpha)^2 \INT\, \bar{\psi}(x)\,\g{\mu}A_{\mu}(x)\,\psi(x)\\
		&+\I\sin(\alpha)\cos(\alpha)\INT\,\bar{\psi}(x)\,\g{\mu}(\I\g{5}\g{0})A_{\mu}(x)\,\psi(\mathcal{P}x)\\
		&+\I\sin(\alpha)\cos(\alpha)\INT\,\bar{\psi}(\mathcal{P}x)\,(\I\g{5}\g{0})\g{\mu}A_{\mu}(x)\,\psi(x)\\
		&-\sin(\alpha)^2 \INT\,\bar{\psi}(\mathcal{P}x)\,(\I\g{5}\g{0})\g{\mu}(\I\g{5}\g{0})A_{\mu}(x)\,\psi(\mathcal{P}x)\\
		&= \cos(\alpha)^2 \INT\, \bar{\psi}(x)\,\g{\mu}A_{\mu}(x)\,\psi(x)\\
		&+\I\sin(\alpha)\cos(\alpha)\INT\,\bar{\psi}(x)\,\g{\mu}(\I\g{5}\g{0})A_{\mu}(x)\,\psi(\mathcal{P}x)\\
		&-\I\sin(\alpha)\cos(\alpha)\INT\,\bar{\psi}(x)\,\mathcal{P}_{\nu}^{\;\mu}\,\g{\nu}(\I\g{5}\g{0})A_{\mu}(\mathcal{P}x)\,\psi(\mathcal{P}x)\\
		&+\sin(\alpha)^2 \INT\,\bar{\psi}(x)\,\g{\nu}\,\mathcal{P}^{\;\mu}_{\nu}\,A_{\mu}(\mathcal{P}x)\,\psi(x)\\
		&=\INT \left[\bar{\psi}(x)\, \g{\mu}(\cos(\alpha)^2 A_{\mu}(x) + \sin(\alpha)^2 A_{\mu}^P(x))\,\psi(x) \right.\\
		&+\left.\I\sin(\alpha)\cos(\alpha)\,\bar{\psi}(x)\, \g{\mu}(A_{\mu}(x) - A_{\mu}^P(x))\,(\I\g{5}\g{0})\,\psi(\mathcal{P}x)\right],
	\end{split}
	\label{eq:actioninteA}
	\end{equation}
	
	\noindent
	where we used that $(\I\g{5}\g{0})\g{\mu}(\I\g{5}\g{0}) = -\g{\nu}\mathcal{P}^{\;\mu}_{\nu}$ and we have changed the variable $x\to\mathcal{P}x$ for the last two terms after the third equality, and used that the Jacobian $\vert\det(\mathcal{P})\vert = 1$. 
	Eq. (\ref{eq:actioninteA}) tells us that for a generic value of $\alpha$, 
	$S_I (\psi^{U_{\PA}^{\alpha}},\bar{\psi}^{U_{\PA}^{\alpha}},A) \neq S_I (\psi,\bar{\psi},A)$ because in general $A_{\mu}^P(x) \equiv \mathcal{P}^{\;\nu}_{\mu} A_{\nu}(x)\neq A_{\mu}(x)$ and therefore $S_I$ is not $U(1)_{\PA}$-invariant.	
	Eq. (\ref{eq:actioninteA}) is similar to Eq. (\ref{eq:actioninte}) for $U(1)_{\PP}$, hence the two consequences below that equation become similar for $U(1)_{\PA}$. 
	In the sense that we can re-obtain the invariance of $S_I$ only for particular values of $\alpha$.  In particular we recognize two cases:
	
	\begin{itemize}
		\item $\alpha = \pi k$ with $k=0,1,2,...$ $\Rightarrow U(1)_{\PA}$ reduces to the group $Z_2 \subset U(1)_{\PA}$ (see Eq. (\ref{eq:u1pa})) and of course $S_F + S_I$ is $Z_2$-invariant.
		\item $\alpha = \pi k+ (\pi/2)$ with $k=0,1,2,...$ $\Rightarrow$ In this case if we perform also a parity transformation of the gauge field $A_{\mu}(x)\to A_{\mu}^{P}(x)$, we can obtain the invariance of the interaction term, namely $S_I (\psi^{U_{\PA}^{\alpha}},\bar{\psi}^{U_{\PA}^{\alpha}},A^P) = S_I (\psi,\bar{\psi},A)$. 
		In fact, as it is clear from Eq. (\ref{eq:u1pa}), $U(1)_{\PA}$ group transformations reduce to Parity $\times\,\g{5}  Z_2 \in U(1)_{\PA}$ transformations, 
		where for Parity $\times\, \g{5} Z_2 $ we mean the transformation: $\psi(x) \to z\,\g{5} (P\psi(x)P^{\dagger})$, with $z\in Z_2$.
	\end{itemize}

	\noindent
	Otherwise a sufficient condition for the $U(1)_{\PA}$ invariance of $S_I (\psi,\bar{\psi},A)$, can be obtained restricting ourself to gauge configurations satisfying the condition in Eq. (\ref{eq:gauge}).

	Regarding the mass term: $m\INT\,\bar{\psi}(x)\psi(x)$, we have 
	
	\begin{equation}
		\begin{split}
		&\INT\,\bar{\psi}(x)^{U_{\PA}^{\alpha}}\psi(x)^{U_{\PA}^{\alpha}}\\
		 &= 
		\cos(\alpha)^2 \INT\,\bar{\psi}(x)\psi(x)\\
		&+\I\sin(\alpha)\cos(\alpha)\INT\,\bar{\psi}(x)(\I\g{5}\g{0})\psi(\mathcal{P}x)\\
		& +\I\sin(\alpha)\cos(\alpha) \INT\,\bar{\psi}(\mathcal{P}x)(\I\g{5}\g{0})\psi(x)\\
		& - \sin(\alpha)^2  
		\INT\,\bar{\psi}(\mathcal{P}x)(\I\g{5}\g{0})(\I\g{5}\g{0})\psi(\mathcal{P}x)\\
		&=\cos(2\alpha)\INT\,\bar{\psi}(x)\psi(x)\\
		&+\I\sin(2\alpha)\INT\,\bar{\psi}(\mathcal{P}x)(\I\g{5}\g{0})\psi(x)\\
		&\neq 	\INT\,\bar{\psi}(x)\psi(x),
		\end{split}
	\end{equation}
	
	\noindent
	which is therefore not invariant under $U(1)_{\PA}$ transformations. The mass term breaks $U(1)_{\PA}$.
	
	\section{Explicit calculation of Eq. (\ref{eq:cpup})}\label{app:cpup} 
	
	Let us start proving the first of Eq. (\ref{eq:cpup}), 
	considering the expression (\ref{eq:uid}) for the $U(1)_{\PP}$ transformations and the expression of $c_r (\bm{p})$ in (\ref{eq:cdprop1}). 
	Therefore we have 
	
	\begin{equation}
	\begin{split}
		c_r (\bm{p})^{U_{\PP}^{\alpha}} &= \int \frac{d^3x}{(2\pi)^{3/2}} u_r(\bm{p})^{\dagger}\psi(x)^{U_{\PP}^{\alpha}}  e^{\I px}\\ 
		&=
		\cos(\alpha)\left[\int \frac{d^3x}{(2\pi)^{3/2}} u_r(\bm{p})^{\dagger}\psi(x) e^{\I px}\right]\\
		&+\I\sin(\alpha)\left[\int \frac{d^3x}{(2\pi)^{3/2}} u_r(\bm{p})^{\dagger}\g{0}\psi(\mathcal{P}x) e^{\I px}\right]\\
		&=\cos(\alpha)\,c_r (\bm{p})\\
		& + \I\sin(\alpha)
		\left[\int \frac{d^3x}{(2\pi)^{3/2}} (\g{0}u_r(\bm{p}))^{\dagger}\psi(x) e^{\I p(\mathcal{P}x)}\right]\\
		&=\cos(\alpha)\,c_r (\bm{p}) + \I\sin(\alpha) \,c_r^P(\bm{p}),
	\end{split}
	\end{equation}
	
	\noindent
	where we have changed the variable $x\to\mathcal{P}x$ and used that $\vert\det(\mathcal{P})\vert = 1$. 
	We have also used the expression of $c_r^P(\bm{p})$ in (\ref{eq:cdprop2}) and Eq. (\ref{eq:uvprop1}). 
	
	For $c_r^P(\bm{p})$ we have  
	
	\begin{equation}
		\begin{split}
		c_r^P (\bm{p})^{U_{\PP}^{\alpha}}&= \int \frac{d^3x}{(2\pi)^{3/2}} u_r(-\bm{p})^{\dagger}\psi(x)^{U_{\PP}^{\alpha}}  e^{\I p(\mathcal{P}x)}\\ 
		&=\cos(\alpha)\left[\int \frac{d^3x}{(2\pi)^{3/2}} u_r(-\bm{p})^{\dagger}\psi(x) e^{\I p(\mathcal{P}x)}\right]\\
		&+\I\sin(\alpha)\left[\int \frac{d^3x}{(2\pi)^{3/2}} u_r(-\bm{p})^{\dagger}\g{0}\psi(\mathcal{P}x) e^{\I (\mathcal{P}x)}\right]\\
		&=\cos(\alpha)\,c_r^P (\bm{p})\\
		& + \I\sin(\alpha)
		\left[\int \frac{d^3x}{(2\pi)^{3/2}} (\g{0}u_r(-\bm{p}))^{\dagger}\psi(x) e^{\I px}\right]\\
		&=\cos(\alpha)\,c_r^P (\bm{p}) + \I\sin(\alpha) \,c_r(\bm{p}),
		\end{split}
	\end{equation}
	
	\noindent
	where we used the same procedure as before. 
	Regarding $d_r (\bm{p})^{\dagger}$ we obtain 
	
	\begin{equation}
	\begin{split}
		(d_r (\bm{p})^{\dagger})^{U_{\PP}^{\alpha}}&= \int \frac{d^3x}{(2\pi)^{3/2}} v_r(\bm{p})^{\dagger}\psi(x)^{U_{\PP}^{\alpha}}  e^{-\I px}\\ 
		&=\cos(\alpha)\left[\int \frac{d^3x}{(2\pi)^{3/2}} v_r(\bm{p})^{\dagger}\psi(x) e^{-\I px}\right]\\
		&+\I\sin(\alpha)\left[\int \frac{d^3x}{(2\pi)^{3/2}} v_r(\bm{p})^{\dagger}\g{0}\psi(\mathcal{P}x) e^{-\I px}\right]\\
			&=\cos(\alpha)\,d_r (\bm{p})^{\dagger}\\
			& + \I\sin(\alpha)
		\left[\int \frac{d^3x}{(2\pi)^{3/2}} (\g{0}v_r(\bm{p}))^{\dagger}\psi(x) e^{-\I p(\mathcal{P}x)}\right]\\
		&=\cos(\alpha)\,d_r (\bm{p})^{\dagger} + \I\sin(\alpha) \,d_r^P(\bm{p})^{\dagger},
	\end{split}
	\end{equation}
	
	\noindent
	and similar for $d_r^P (\bm{p})^{\dagger}$ we have
	
		\begin{equation}
	\begin{split}
	(d_r^P (\bm{p})^{\dagger})^{U_{\PP}^{\alpha}} &= \int \frac{d^3x}{(2\pi)^{3/2}} (-v_r(-\bm{p}))^{\dagger}\psi(x)^{U_{\PP}^{\alpha}}  e^{-\I p(\mathcal{P}x)}\\ 
	&=\cos(\alpha)\left[\int \frac{d^3x}{(2\pi)^{3/2}} (-v_r(-\bm{p}))^{\dagger}\psi(x) e^{-\I p(\mathcal{P}x)}\right]\\
	&-\I\sin(\alpha)\left[\int \frac{d^3x}{(2\pi)^{3/2}} v_r(-\bm{p})^{\dagger}\g{0}\psi(\mathcal{P}x) e^{-\I p(\mathcal{P}x)}\right]\\
	&=\cos(\alpha)\,d_r^P (\bm{p})^{\dagger} \\
	&- \I\sin(\alpha)
	\left[\int \frac{d^3x}{(2\pi)^{3/2}} (\g{0}v_r(-\bm{p}))^{\dagger}\psi(x) e^{-\I px}\right]\\
	&=\cos(\alpha)\,d_r^P (\bm{p})^{\dagger} + \I\sin(\alpha) \,d_r(\bm{p})^{\dagger},
	\end{split}
	\end{equation}
	
		\noindent
	where we have changed the variable $x\to\mathcal{P}x$ and used that $\vert\det(\mathcal{P})\vert = 1$. 
	We have also used the expressions of $d_r(\bm{p})^{\dagger}$ and $d_r^P(\bm{p})^{\dagger}$ in (\ref{eq:cdprop1}) and (\ref{eq:cdprop2}) respectively, and Eq. (\ref{eq:uvprop1}).

	\section{Explicit calculation of Eqs. (\ref{eq:csu2cs}) and (\ref{eq:dsu2cs})}\label{app:cdsu2cs} 
	
	We have seen in section \ref{sec:sunham} that we can write $u_{r}(\bm{p})^{\dagger}\psi(x) = U_r (\bm{p})^{\dagger}\spiu\Psi(x)$ and $v_{r}(\bm{p})^{\dagger}\psi(x) = V_r (\bm{p})^{\dagger}\spiu\Psi(x)$, 
	where $U_r (\bm{p})$ and $V_r (\bm{p})$ are defined at the beginning of that section. 
	The properties of $u_r (\bm{p})$ and $v_r (\bm{p})$ (see Eq. (\ref{eq:uv})), given in Eqs. (\ref{eq:uvprop1}) and (\ref{eq:uvprop2}), valid for $m=0$, can be translated for $U_r (\bm{p})$ and $V_r (\bm{p})$ as well, 
	and we obtain
	
	\begin{equation}
		\begin{split}
		&(\mathds{1}\otimes\g{0})U_{r}(\bm{p}) = U_{r}(-\bm{p}),\\ &(\mathds{1}\otimes\g{0})V_{r}(\bm{p}) = -V_{r}(-\bm{p}),\\
		&(\mathds{1}\otimes\g{5})U_{r}(\bm{p}) = h_r U_{r}(\bm{p}),\\ &(\mathds{1}\otimes\g{5})V_{r}(\bm{p}) = h_{r\oplus 1} V_{r}(\bm{p}),\\
		\end{split}
		\label{eq:UV}
	\end{equation}
	
	\noindent
	where 3rd and 4th equations are valid for $m= 0$, where $\g{5}$ coincides with the helicity operator $\bm{\sigma}\cdot\bm{p}/\vert\bm{p}\vert$. This implies that for opposite values $- \bm{p }$, 3rd and 4th equations of (\ref{eq:UV}) become 
	
	\begin{equation}
		\begin{split}
	&(\mathds{1}\otimes\g{5})U_{r}(-\bm{p}) = -h_r U_{r}(-\bm{p}),\\ &(\mathds{1}\otimes\g{5})V_{r}(-\bm{p}) = -h_{r\oplus 1} V_{r}(-\bm{p}).
		\end{split}
		\label{eq:UV2}
	\end{equation}
	
	Using such notation 
	and defining for brevity $\ggu=\spiu$
	, we can rewrite $c_r (\bm{p})$, $c_r^P (\bm{p})$, $d_r (\bm{p})^{\dagger}$ and $d_r^P (\bm{p})^{\dagger}$ of Eqs. (\ref{eq:cdprop1}) and (\ref{eq:cdprop2}) as 
	
	\begin{equation}
		\begin{split}
		 c_r (\bm{p}) &= \int \frac{d^3 x}{(2\pi)^{3/2}} U_r (\bm{p})^{\dagger}\ggu\Psi(x) e^{\I px},\\
		 c_r^P (\bm{p}) &= \int \frac{d^3 x}{(2\pi)^{3/2}} U_r (-\bm{p})^{\dagger}\ggu \Psi(x) e^{\I p(\mathcal{P}x)}\\
		&= \int \frac{d^3 x}{(2\pi)^{3/2}} U_r (-\bm{p})^{\dagger}\ggu \Psi(\mathcal{P}x) e^{\I px},\\
		 d_r (\bm{p})^{\dagger} &= \int \frac{d^3 x}{(2\pi)^{3/2}} V_r (\bm{p})^{\dagger}\ggu \Psi(x) e^{-\I px},\\
		 d_r^P (\bm{p})^{\dagger} &= \int \frac{d^3 x}{(2\pi)^{3/2}} (-V_r (-\bm{p}))^{\dagger}\ggu \Psi(x) e^{-\I p(\mathcal{P}x)}\\
		& = \int \frac{d^3 x}{(2\pi)^{3/2}} (-V_r (-\bm{p}))^{\dagger}\ggu \Psi(\mathcal{P}x) e^{-\I px},\\
		\end{split}
		\label{eq:step0}
	\end{equation}
	
	\noindent
	where in $c_r^P (\bm{p})$ and $d_r^P (\bm{p})^{\dagger}$, we have changed the variable $x\to \mathcal{P}x$ and used that the Jacobian is $\vert\det(\mathcal{P})\vert = 1$. 
	At this point the $\sun$ transformations are given by 
	
	\begin{widetext}
	\begin{equation}
		\begin{split}
		&c_r (\bm{p})^{\sutr} = \int \frac{d^3 x}{(2\pi)^{3/2}} U_r (\bm{p})^{\dagger}\ggu \Psi(x)^{\sutr} e^{\I px}\\ 
		&= \cos(\alpha)\,c_r (\bm{p}) + \I\sin(\alpha)\sum_{i=1}^3 e_i 
		\int \frac{d^3 x}{(2\pi)^{3/2}} U_r (\bm{p})^{\dagger}\ggu \Sigma_i^{\mathcal{P}}\Psi(x) e^{\I px},\\
		&c_r^P (\bm{p})^{\sutr} = \int \frac{d^3 x}{(2\pi)^{3/2}} U_r (-\bm{p})^{\dagger}\ggu \Psi(x)^{\sutr} e^{\I p(\mathcal{P}x)} \\
		&= \cos(\alpha)\,c_r^P (\bm{p}) + \I\sin(\alpha)\sum_{i=1}^3 e_i 
		\int \frac{d^3 x}{(2\pi)^{3/2}} U_r (-\bm{p})^{\dagger}\ggu\Sigma_i^{\mathcal{P}}\Psi(x) e^{\I p(\mathcal{P}x)},\\	
		&(d_r (\bm{p})^{\dagger} )^{\sutr}= \int \frac{d^3 x}{(2\pi)^{3/2}} V_r (\bm{p})^{\dagger}\ggu\Psi(x)^{\sutr} e^{-\I px} \\
		&= \cos(\alpha)\,d_r (\bm{p})^{\dagger} + \I\sin(\alpha)\sum_{i=1}^3 e_i 
		\int \frac{d^3 x}{(2\pi)^{3/2}} V_r (\bm{p})^{\dagger}\ggu \Sigma_i^{\mathcal{P}}\Psi(x) e^{-\I px},\\
		&(d_r^P (\bm{p})^{\dagger})^{\sutr} = \int \frac{d^3 x}{(2\pi)^{3/2}} (-V_r (-\bm{p}))^{\dagger}\ggu\Psi(x)^{\sutr} e^{-\I p(\mathcal{P}x)} \\
		&= \cos(\alpha)\,d_r^P (\bm{p})^{\dagger} + \I\sin(\alpha)\sum_{i=1}^3 e_i 
		\int \frac{d^3 x}{(2\pi)^{3/2}} (-V_r (-\bm{p}))^{\dagger}\ggu\Sigma_i^{\mathcal{P}}\Psi(x) e^{-\I p(\mathcal{P}x)},\\		
		\end{split}
		\label{eq:step1}
	\end{equation}
	\end{widetext}

	\noindent
	where we used that $ \Psi(x)^{\sutr}  = \sutr\Psi(x)$ and $\sutr$ is given in Eq. (\ref{eq:sutr}). 
	From $c_r (\bm{p})$ and $c_r^P (\bm{p})$ is evident that we need to know the terms $U_r (\pm\bm{p})^{\dagger}\spiu\Sigma_i^{\mathcal{P}}\Psi(x)$, where $\Sigma_i^{\mathcal{P}} =\{\sigma^3 \otimes \g{0},\sigma^3 \otimes \I\g{5}\g{0},-\mathds{1}\otimes\g{5}\}$. 
	At first we notice that by definition $(\sigma^3 \otimes \mathds{1})\Psi(x) = \Psi(\mathcal{P}x)$ and 
	then from Eqs. (\ref{eq:UV}) and (\ref{eq:UV2}), we get 
	
	\begin{widetext}
	\begin{equation}
		\begin{split}
		 &U_r (\pm\bm{p})^{\dagger}\ggu \Sigma_1^{\mathcal{P}}\Psi(x) = 
		 ((\mathds{1}\otimes\g{0})U_r (\pm\bm{p}))^{\dagger}\ggu  (\sigma^3 \otimes \mathds{1})\Psi(x) 
		 = U_r (\mp\bm{p})^{\dagger}\ggu \Psi(\mathcal{P}x),\\
		 &U_r (\pm\bm{p})^{\dagger}\ggu \Sigma_2^{\mathcal{P}}\Psi(x) = 
		 ((\mathds{1}\otimes\I\g{5}\g{0})U_r (\pm\bm{p}))^{\dagger}\ggu  (\sigma^3 \otimes \mathds{1})\Psi(x) 
		 = \pm\I h_r U_r (\mp\bm{p})^{\dagger}\ggu \Psi(\mathcal{P}x),\\
		 &U_r (\pm\bm{p})^{\dagger}\ggu \Sigma_3^{\mathcal{P}}\Psi(x) = 
		 ((-\mathds{1}\otimes\g{5})U_r (\pm\bm{p}))^{\dagger}\ggu \Psi(x) 
		 =\mp h_r U_r (\pm\bm{p})^{\dagger}\ggu \Psi(x) ,
		\end{split}
		\label{eq:cpart}
	\end{equation}
	\end{widetext}

	\noindent
	where we have decomposed $\sigma^3 \otimes l_i = (\mathds{1}\otimes l_i)(\sigma^3 \otimes \mathds{1})$, 
		with $l_i =\{\g{0},\I\g{5}\g{0}\}$ and used that $[\mathds{1}\otimes l_i,\Gamma]=0$ for $i=1,2$. 
		Now we can plug (\ref{eq:cpart}) in Eq. (\ref{eq:step1}), having in mind (\ref{eq:step0}), 
		and finally we obtain the right sides of Eq. (\ref{eq:csu2cs}). 
		
		Regarding the terms $V_r (\bm{p})^{\dagger}\spiu\Psi(x)^{\sutr} $ for $d_r (\bm{p})^{\dagger}$ and $d_r^P (\bm{p})^{\dagger}$, they can be rewritten using Eqs. (\ref{eq:UV}) and (\ref{eq:UV2}) as 
		\begin{widetext}
		\begin{equation}
		\begin{split}
		&V_r (\pm\bm{p})^{\dagger}\ggu \Sigma_1^{\mathcal{P}}\Psi(x) = 
		((\mathds{1}\otimes\g{0})V_r (\pm\bm{p}))^{\dagger}\ggu  (\sigma^3 \otimes \mathds{1})\Psi(x) 
		= -V_r (\mp\bm{p})^{\dagger}\ggu \Psi(\mathcal{P}x),\\
		&V_r (\pm\bm{p})^{\dagger}\ggu \Sigma_2^{\mathcal{P}}\Psi(x) = 
		((\mathds{1}\otimes\I\g{5}\g{0})V_r (\pm\bm{p}))^{\dagger}\ggu  (\sigma^3 \otimes \mathds{1})\Psi(x) 
		= \mp\I h_{r\oplus 1} V_r (\mp\bm{p})^{\dagger}\ggu \Psi(\mathcal{P}x),\\
		&V_r (\pm\bm{p})^{\dagger}\ggu \Sigma_3^{\mathcal{P}}\Psi(x) = 
		((-\mathds{1}\otimes\g{5})V_r (\pm\bm{p}))^{\dagger}\ggu \Psi(x) 
		=\mp h_{r\oplus 1} V_r (\pm\bm{p})^{\dagger}\ggu \Psi(x) ,
		\end{split}
		\label{eq:dpart}
		\end{equation}
		\end{widetext}
	
		\noindent
		where we used the same procedure as in Eq. (\ref{eq:cpart}). 
		Now we plug (\ref{eq:dpart}) in Eq. (\ref{eq:step1}) and keeping in mind the expression (\ref{eq:step0}), we obtain the right sides of Eq. (\ref{eq:dsu2cs}). 
		This ends our calculation.
	
	\bibliographystyle{unsrtnat}
	\bibliography{su2cs_catillo_2021_2}
	
\end{document}